\documentclass[footinbib,twocolumn,amsmath,amstex,amssymb,mathfonts,superscriptaddress,prl]{revtex4-1}
\usepackage{graphicx}

\usepackage{gbt7714}

\usepackage{caption}
\usepackage[normalem]{ulem}
\usepackage{dcolumn}

\usepackage{float}
\newcommand{\bigO}{\mathcal{O}}

\renewcommand{\raggedright}{\rightskip=0pt \leftskip=0pt plus 0cm}
\usepackage{booktabs}

\usepackage{algorithm,algcompatible}

\algnewcommand\INPUT{\item[\textbf{Input:}]}%
\algnewcommand\OUTPUT{\item[\textbf{Output:}]}%
\algnewcommand\PREPARATION{\item[\textbf{Preparation:}]}%
\algnewcommand\TRAINING{\item[\textbf{\underline{Training Stage:}}]}%
\algnewcommand\INFERENCE{\item[\textbf{\underline{Inference Stage:}}]}%

\usepackage{epstopdf}

\usepackage[usenames, dvipsnames]{color}
\usepackage{bm}
\usepackage{verbatim}

\usepackage{physics}
\usepackage{amsmath}
\usepackage{amssymb}
\usepackage{amsthm}
\usepackage{amsfonts}
\usepackage{soul}
\usepackage{float}
\usepackage[caption=true]{subfig}
\usepackage{bbm}
\usepackage{bbm}
\begin{document}
\def\qv{\vec{q}}
\def\red{\textcolor{red}}
\def\blue{\textcolor{blue}}
\def\apricot{\textcolor{Apricot}}

\def\GJ{\textcolor{blue}}
\def\YH{\textcolor{Orange}}
\def\TT{\textcolor{ForestGreen}}

\newcommand{\ad}[1]{\text{ad}_{S_{#1}(t)}}

\title{Predicting quantum many-body dynamics with transferable neural networks}

\author{Ze-Wang Zhang}
\affiliation{School of Physics, Sun Yat-sen University, Guangzhou 510275, China. Telephone: 15602298593}
\author{Shuo Yang}
\affiliation{State Key Laboratory of Low-Dimensional Quantum Physics and Department of Physics, Tsinghua University, Beijing 100084, China}
\author{Yi-Hang Wu}
\affiliation{School of Physics, Sun Yat-sen University, Guangzhou 510275, China}
\author{Chen-Xi Liu}
\affiliation{School of Physics, Sun Yat-sen University, Guangzhou 510275, China}
\author{Yi-Min Han}
\affiliation{School of Physics, Sun Yat-sen University, Guangzhou 510275, China}
\author{Ching-Hua Lee}
\affiliation{Department of Physics, National University of Singapore, 117542, Singapore}
\affiliation{Institute of High Performance Computing, 138632, Singapore}
\author{Zheng Sun}
\affiliation{School of Physics, Sun Yat-sen University, Guangzhou 510275, China}
\author{Guang-Jie Li}
\affiliation{School of Physics, Sun Yat-sen University, Guangzhou 510275, China}
\author{Xiao Zhang}
\email{zhangxiao@mail.sysu.edu.cn}
\affiliation{School of Physics, Sun Yat-sen University, Guangzhou 510275, China}
\date{\today}
\begin{abstract}
Machine learning (ML) architectures such as convolutional neural networks (CNNs) have garnered considerable recent attention in the study of quantum many-body systems. However, advanced ML approaches such as transfer learning have seldom been applied to such contexts. Here we demonstrate that a simple recurrent unit (SRU) based efficient and transferable sequence learning framework is capable of learning and accurately predicting the time evolution of one-dimensional (1D) Ising model with simultaneous transverse and parallel magnetic fields, as quantitatively corroborated by relative entropy measurements and magnetization between the predicted and exact state distributions. At a cost of constant computational complexity, a larger many-body state evolution was predicted in an autoregressive way from just one initial state, without any guidance or knowledge of any Hamiltonian.  Our work paves the way for future applications of advanced ML methods in quantum many-body dynamics only with knowledge from a smaller system. 

\end{abstract}

\maketitle 
\section{Introduction}
Machine learning (ML) approaches, particularly neural networks (NNs), have achieved great success in solving real-world industrial and social problems~\citep{bengio2013representation}, such as image recognition\citep{krizhevsky2012imagenet}, high level image synthesis and style transfer\citep{gatys2016image}, human-like raw speech generator\cite{van2016wavenet}, producing original melodious MIDI notes\citep{sun2018composing}, neural machine translation~\citep{wu2016google}. Inspired by its widespread applicability, ML was soon adopted by condensed matter physicists in the modeling of quantum many-body behavior and phase transition discovery~\citep{van2017learning, cai2018approximating, torlai2016learning, torlai2017neural, deng2017quantum, gray2018machine, ch2017machine, broecker2017machine, amin2018quantum, zhang2017quantum, carleo2017solving}. Compared to so many advances in computer vision\citep{voulodimos2018deep}, speech processing\citep{deng2013new}, and natural language processing~\citep{young2018recent}, it is natural to ask if recent progress in these more sophisticated ML architectures can benefit or even revolutionize the modeling of quantum systems. For instance, can quantum many-body dynamics be ``learned" through transferable learning\cite{pan2009survey,torrey2010transfer}.

\begin{figure*}
\subfloat[]{\includegraphics[width= 3.0in]{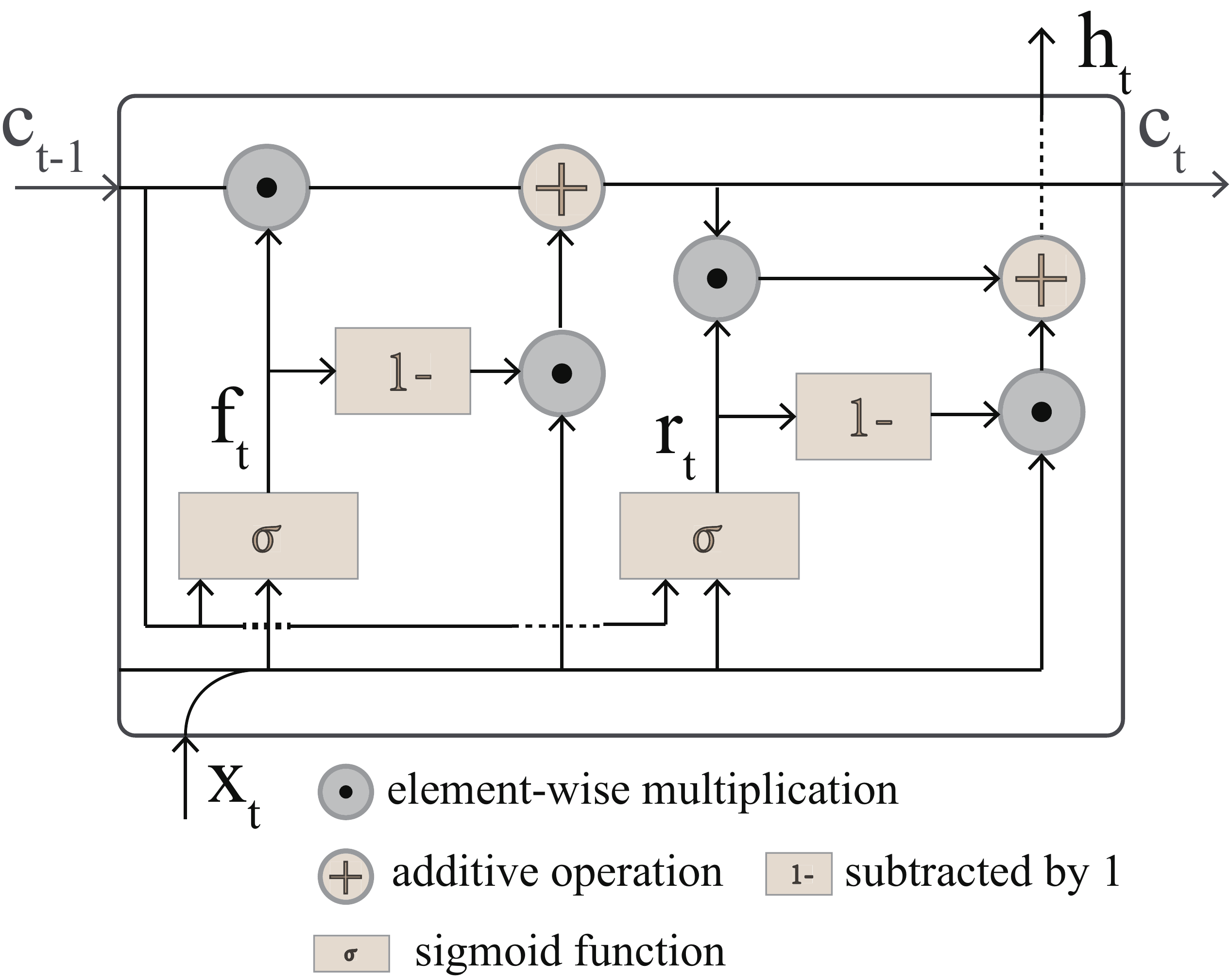}\label{sru}}
\qquad
\subfloat[]{\includegraphics[width= 3.0in]{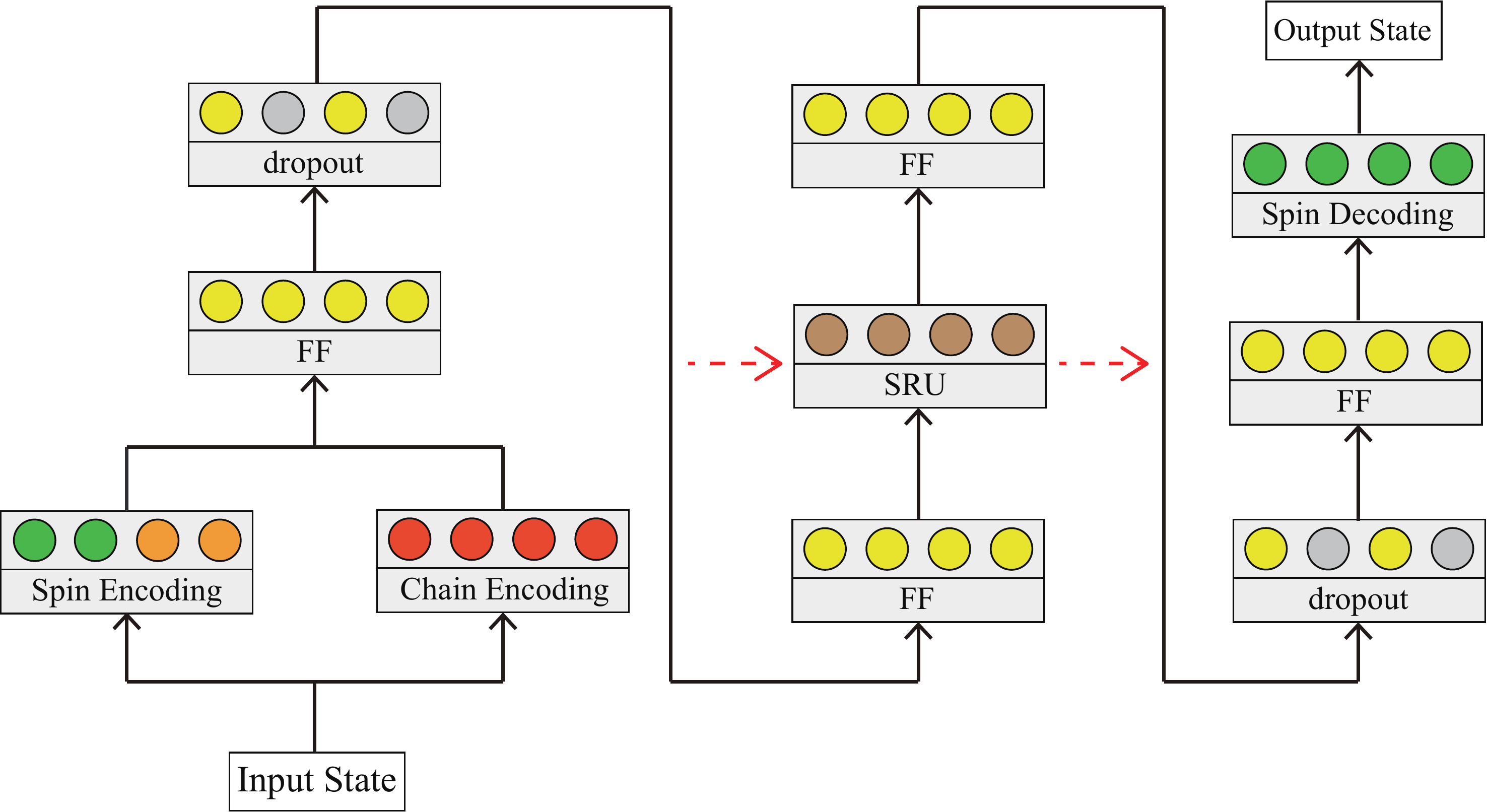}\label{framework}} 
\hfill
\subfloat[]{\includegraphics[width= 3.0in]{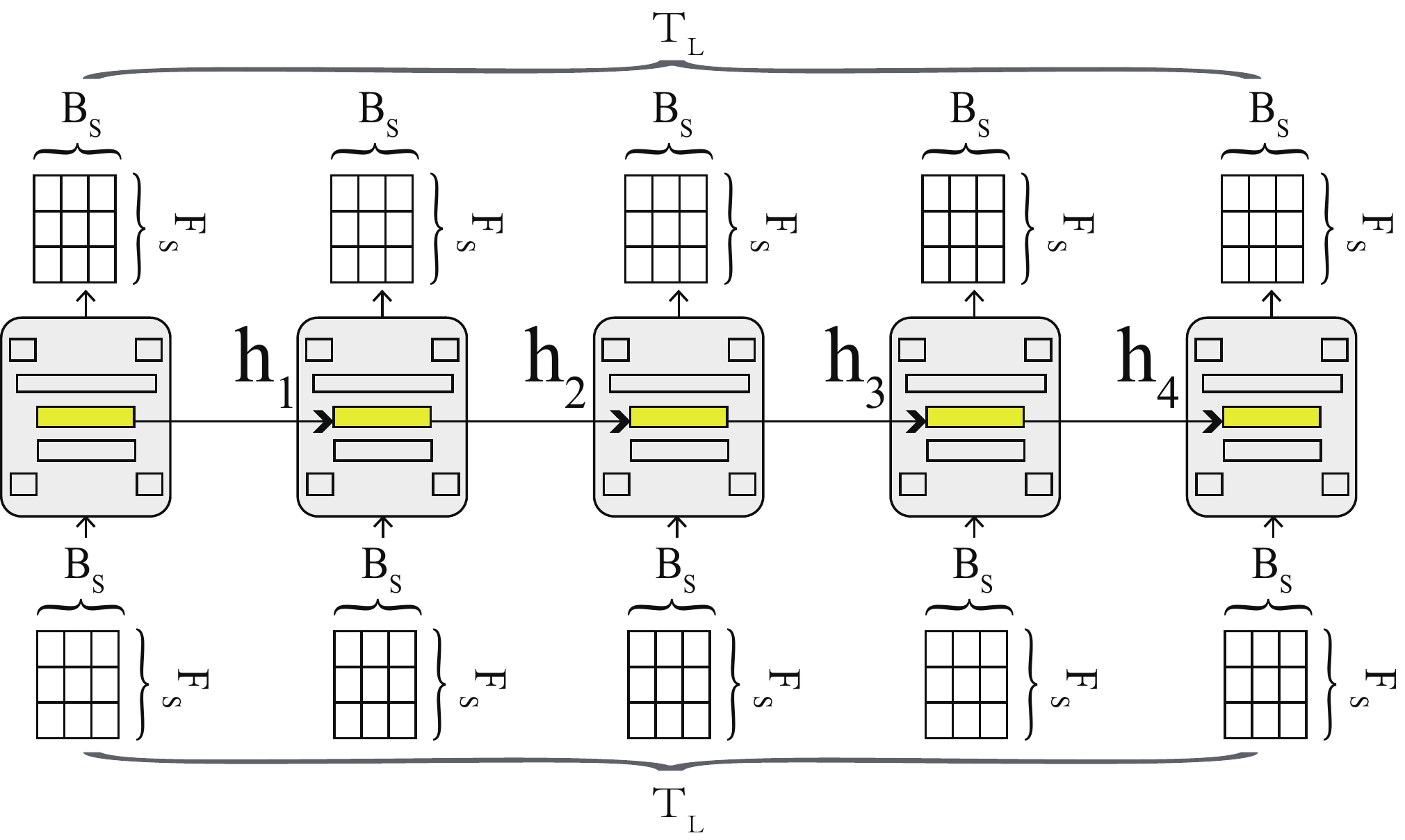}\label{framework_train}}
\qquad
\subfloat[]{\includegraphics[width= 3.0in]{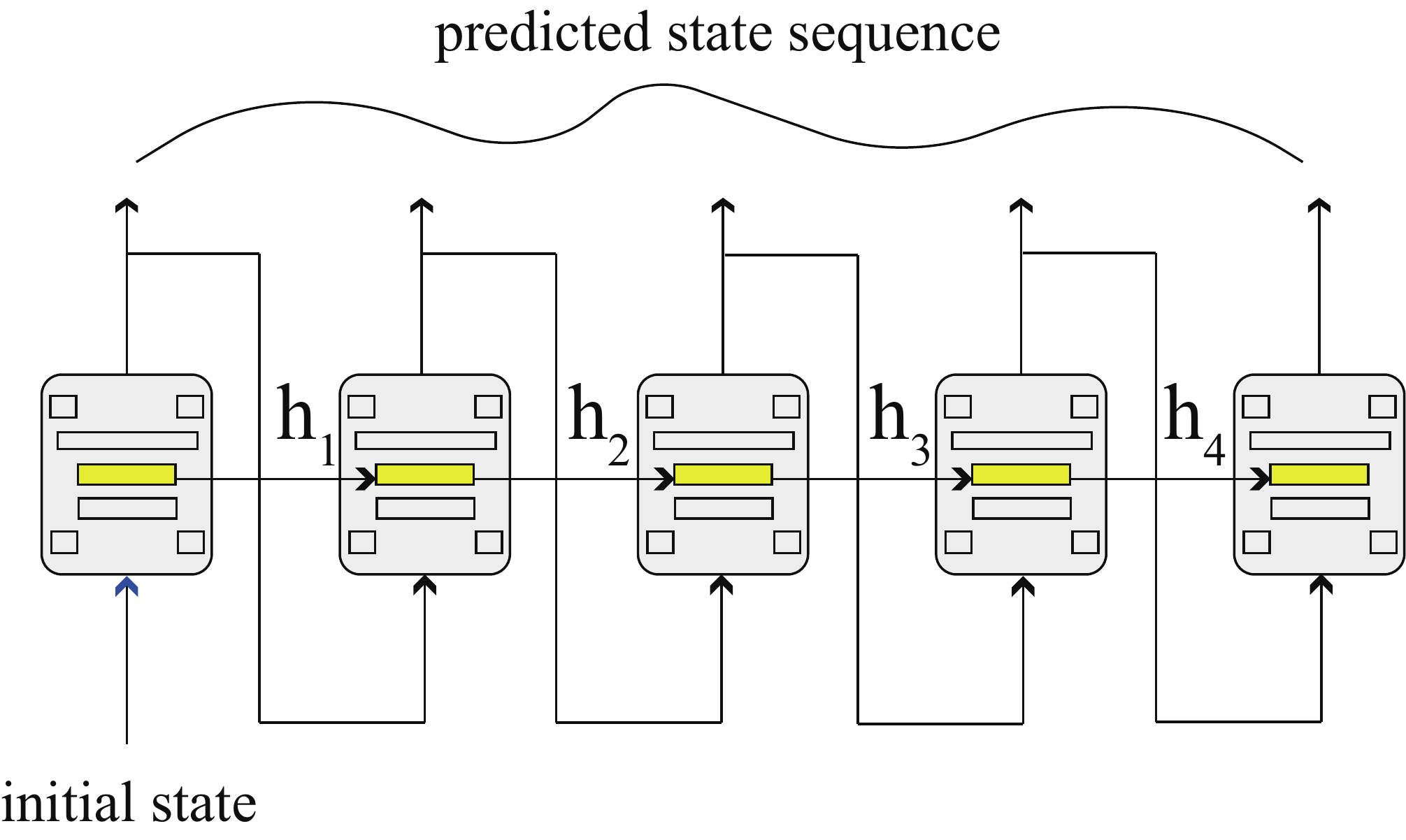}\label{framework_inference}}

\caption{\protect\subref{sru} Details of a vanilla SRU cell with forget gate $f_t$, skip gate $r_t$, memory cell $c_t$ and weighted connections from previous memory cell $c_{t-1}$ to both forget gate $f_t$ and skip gate $r_t$. $X_t$ is the input and $h_t$ is the output hidden state of SRU cell. The forget gate decides what to forget from previous memory cell $c_{t-1}$ and what redundant information to drop when adapting to other systems. The skip gate, along with current memory cell $c_t$, decides what to skip from input and what to output as $h_t$. $\sigma(x)=\frac{e^x}{e^x+1}$ is the activation function. \protect\subref{framework} Proposed architecture in our paper, which we take as a block composed of a spin encoding layer, a chain encoding layer, SRU layers and a spin decoding layer at each timestep. Spin encoding, chain encoding and spin decoding are all feed-forward layers (labeled as ``FF''). \protect\subref{framework_train} Traing the network in a teacher forcing mode\citep{williams1989learning}, we take the current wave functions as input and next state as the ground label. The ground label sequence is one timestep off the input sequence. \protect\subref{framework_inference} Autoregressive procedure of generating new quantum states, given an initial state at the beginning, each time-evolved quantum state is predicted from our proposed block by SRU's cell memory information (shown as arrows in the middle) and previous predicted state.}
\label{figframework}
\end{figure*}

Thus, the main objective of this work is to demonstrate the novel application of NNs in the transferable learning and prediction of the evolution of a many-body wavefunction, an otherwise computationally intensive task that has not been solved by generative models. Focusing on static problems, it is proven that deep NNs like restricted Boltzmann Machine (RBM) can represent most physical states\citep{gao2017efficient}, and a recent work based on very deep and large CNNs shows the ability to circumvent the need for Markov Chain sampling on two-dimensional interacting spin model of larger systems\citep{sharir2019deep}. Lately, physical properties of spin Hamiltonians are reproduced by deep Boltzmann Machine (DBM), as an alternative to the standard path integral\citep{carleo2018constructing}. Our approach is fundamentally in contrast with conventional approaches in computing many-body dynamics: instead of evolving the wavefunction explicitly with the Hamiltonian, which becomes prohibitively slow and impractical as the number of spin variables increases, we directly predict the dynamical wavefunction from the initial state by propagating it with an efficient and transferable framework based on unified spin encoding, chain encoding and SRU\citep{lei2018simple} module. With the same level of parallelism as feed-forward CNNs and scalable context-dependent capacity of recurrent connections, our proposed framework are naturally suited for learning many-body systems with unified parameters, although they have never been harnessed for exact quantum state evolution, in our scenario, a 1D Ising model with both parallel and transverse magnetic field.

Inspired by end-to-end training\citep{graves2014towards} and domain adaptation\citep{pan2010domain,glorot2011domain}, we specialize to the many-body dynamics of a 1D Ising chain with transverse and parallel magnetic fields. Comparison with exact conventionally computed results with up to seven spins reveals high predictive accuracy, as quantified by the relative entropy as well as magnetization. Indeed, our SRU-propagated wavefunction shows a strong grasp of the periodicity in the time evolution, despite being unaware of the Hamiltonian that sets the energy (inverse periodicity) scale. Encouraged by circumventing the problem of exponential computational complexity through unified encoding mechanisms and parallel recurrent connections, we hope that such encouraging results from our pioneering transferable learning appoach will inspire further applications of transferable learning methods to build a shared model suited for quantum systems with vast spin variables.

\section{Dynamics on a 1D Ising chain}
We consider a 1D Ising spin chain composed of $N$ spin variables with local transverse ($g$) and parallel ($h$) magnetic fields, described by the Hamiltonian
\begin{eqnarray}
H=-\sum_{i}^{N-1} \sigma_{i}^{z}\otimes \sigma_{i+1}^{z}-h \sum_{i}^{N}\sigma_{i}^{z} -g \sum_{i}^{N} \sigma_{i}^{x}.
\end{eqnarray}
where $\sigma^{x}$ and $\sigma^{z}$ denotes the Pauli matrices, and $i$ denotes the spin variable. When the magnetic field is parallel ($g=0$) or transverse ($h=0$), the Hamiltonian is exactly solvable. However, when $g \neq 0$ and $h\neq 0$, the dynamics of $N$ spins must be numerically computed in the $2^N$-dimensional many-body Hilbert space spanned by direct product states $\Psi$ of single-spin wavefunctions $\psi_i$:
\begin{eqnarray}
\Psi= \prod_{i}^{N} \otimes \,\psi_{i}=\prod_{i}^{N}\otimes 
\begin{pmatrix} \phi_{i}^{\uparrow} \\ \phi_{i}^{\downarrow} \end{pmatrix}, \ \ \ \ \dim \Psi = 2^{N}.
\end{eqnarray}
Wavefunction dynamics can be exactly computed through unitary time evolution of the Hamiltonian
\begin{eqnarray}
\ket{\Psi(t)}=\exp\qty(-\mathrm{i}\frac{H}{\hbar} t) \ket{\Psi(0)}= V \exp\qty(-\mathrm{i}\frac{E}{\hbar} t)V^{-1} \ket{\Psi(0)}, \nonumber \\
\label{eq:1}
\end{eqnarray}
where $E=V^{-1} H V$ is the diagonal eigenenergy matrix.

The $2^N$-dimensional $N$-body wave function $\ket{\Psi(t)}$ quickly becomes expensive to compute as $N$ increases. We propose a ML approach with spin encoding layer, chain encoding layer, SRU layers, and spin decoding layer, which instead attemps to predict its time evolution based on prior knowledge of the time evolution behavior of known training states. This training (learning) only has to be performed once for the relatively inexpensive prediction of any number of initial states. Importantly, the training and prediction process captures solely the intrinsic evolution patterns of the wavefunctions, and does not involve any explicit knowledge about the Hamiltonian. From the ML perspective, this dynamical state evolution problem can be regarded as a straightforward sequence generation problem~\citep{sutskever2014sequence}. Moreover, as we shall explain, our SRU-based framework is transferable.

\section{The transferable NN approach}

We next outline the broad principles behind our NN approach of predicting quantum state evolution, with details in \citep{SOM}. Here we choose a NN composed of a spin encoding layer, a chain encoding layer, SRU layers and a spin decoding layer (Fig. \ref{framework}). The vanilla SRU NN with peephole connections (Fig.\ref{sru}) substitutes inherent matrix multiplication with parallelizable element-wise multiplication operations ($\odot$ in Fig.~\ref{sru}) associated with $c_{t-1}$, hence the calculation of $f_t$ doesn't have to wait until the whole $c_{t-1}$ is updated. With the help of spin encoding and decoding layers, the amount of trained parameters is fixed, and thus the complexity has an upper bound instead of increase exponentially.

\begin{figure}
\includegraphics[width= .9\linewidth]{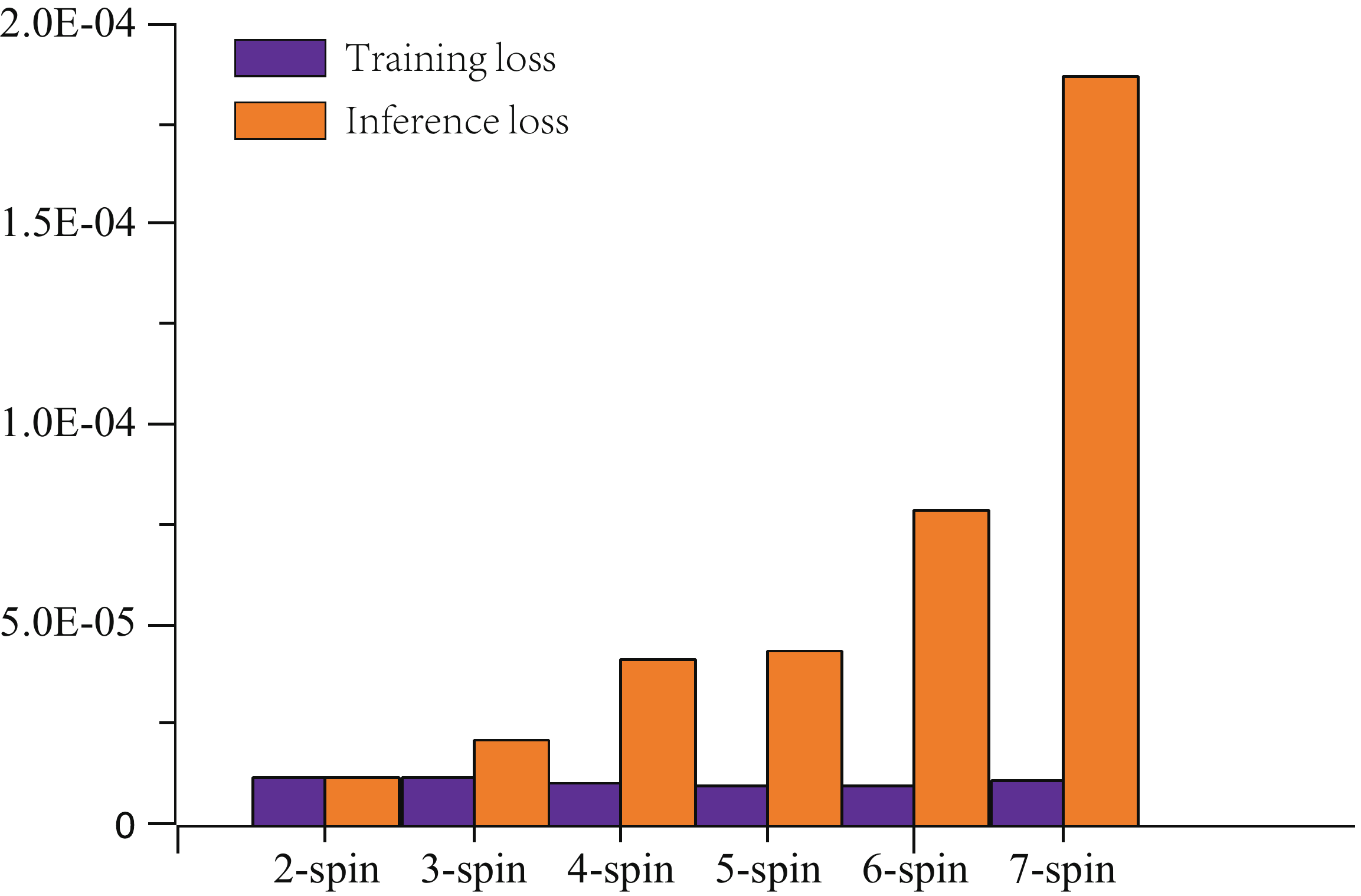}
\caption{Training loss and inference loss of different 1D Ising systems. Inference loss represents the average loss of wavefunction evolution in $100$ timesteps, the maximum inference loss is within 2.0E-4.}
\label{loss}
\end{figure}

\begin{figure}
\includegraphics[width= 1.0\linewidth]{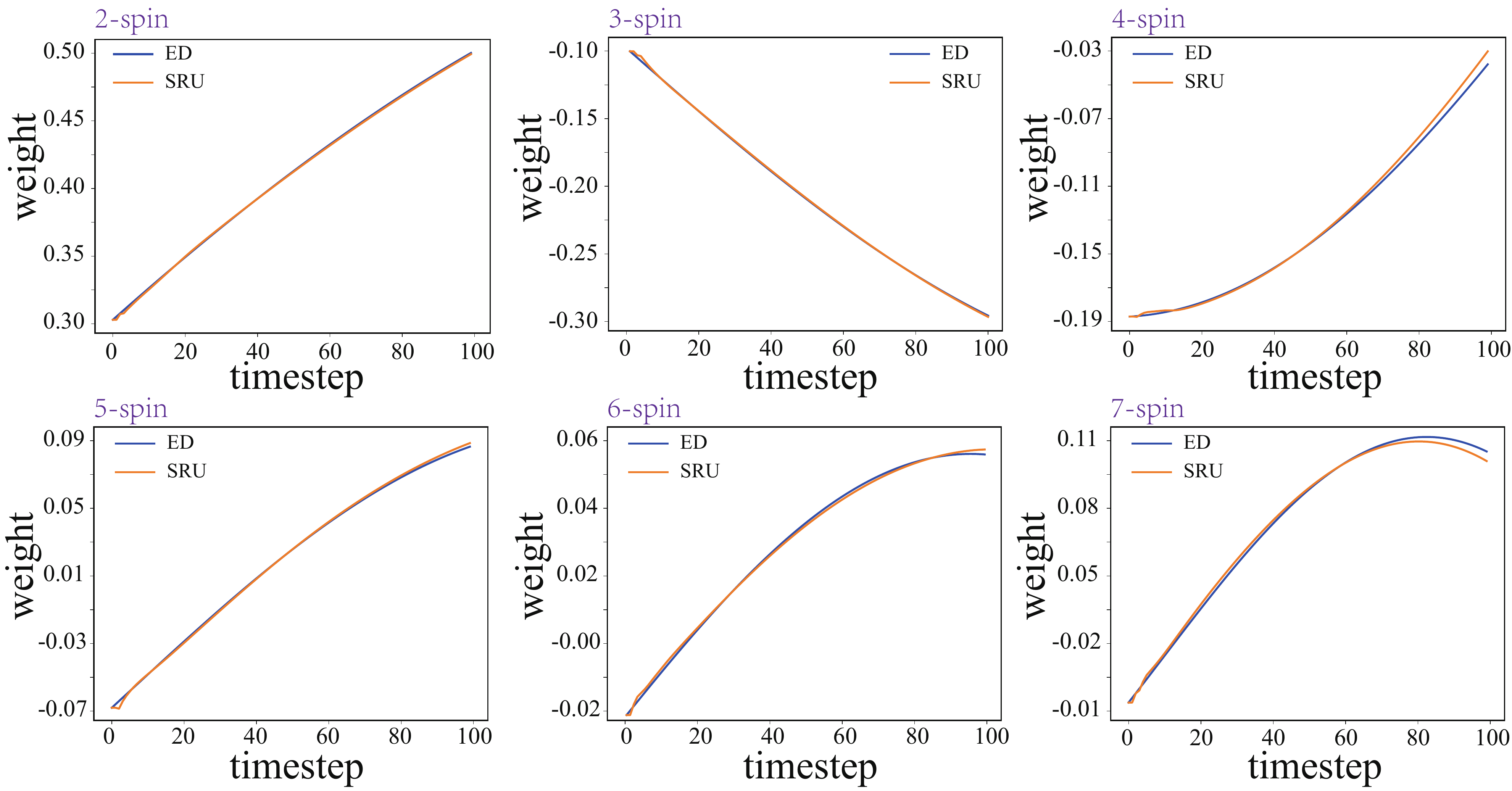}
\caption{Output (SRU-based) and target (ED-based) wavefunction magnitude (y-axis) for $2$-spin to $7$-spin systems, whose initial states are given in \citep{SOM}. We plot the curves of the total $100$ timesteps.}
\label{figresult}
\end{figure}

Our procedure occurs in two main stages: the training stage and the inference stage. In the training stage, we first ``train" or optimize the weight parameters of our SRU-based framework by feeding it with a large number of training sequences, which are the time-evolved wavefunction data of $10^4$ randomly chosen initial $2$-spin to $7$-spin state sequences sampled over 500 timesteps, obtained via conventional exact diagonalization (ED). The SRU-based framework is fully optimized by Adam optimization algorithm~\citep{kingma2014adam} to minimize the mean squared error between the ED-evolved and SRU-evolved states at all timesteps in a mini-batch \citep{SOM}.

Following the training stage is the inference stage, when the SRU-based framework is ready for predicting the evolution of arbitrarily given initial states. As sketched in Fig.~\ref{framework_inference}, the initial many-body state $\ket{\Psi(t=0)}$ enters the leftmost block at $t=0$, then processed by a spin encoding layer, a chain encoding layer, and two fully-connected layers, and its output is propagated as input state to the next block with hidden layers $h_t$. The output of each block denotes a new quantum state at a certain timestep. The combination of memory cell $c_t$ and hidden output $h_t$ serves to implement effective context-dependent behaviors. As illustrated in Fig.~\ref{figframework}(a) and further elaborated in \citep{SOM}, context-dependent information kept in memory cell $c_t$ is modified by its previous value $c_{t-1}$, new input $x_t$ interacted with forget gate and skip gate at that timestep, as well as ``hidden'' information on $h_{t-1}$ from the previous SRU cell. Based on the already optimized SRU-based framework, the final predicted quantum state as a function of time would be generated from one fully-connected layer and the spin decoding layer as shown in Fig.~\ref{framework}.

\section{Comparison between exact and SRU-based evolutions}

We report very encouraging agreements between wavefunctions evolved by $e^{-iHt/\hbar}$ as computed by ED, and wavefunction evolutions as predicted by our SRU-based framework. As for the 1D Ising model, we set the local transverse magnetic field $g$ to be $-1.05$, parallel magnetic field $h$ to be $0.5$ and $\Delta t$, the time interval to be $0.002$, and keep this setting constant for all computation. We find that the maximum energy eigenvalue is about $0.1\gg0.002$, proving that the time interval we choose is small enough. The number of spin variables studied ($2$ to $7$) decides the cost of exactly computing the $10^4$ different time evolutions over 0.2 second (100 timesteps) prior to training the network, since the time complexity of ED method is $\bigO (2^n)$. The training and inference loss of different systems is shown in Fig.~\ref{loss}. 

\begin{figure}
\subfloat[]{\includegraphics[width= .9\linewidth]{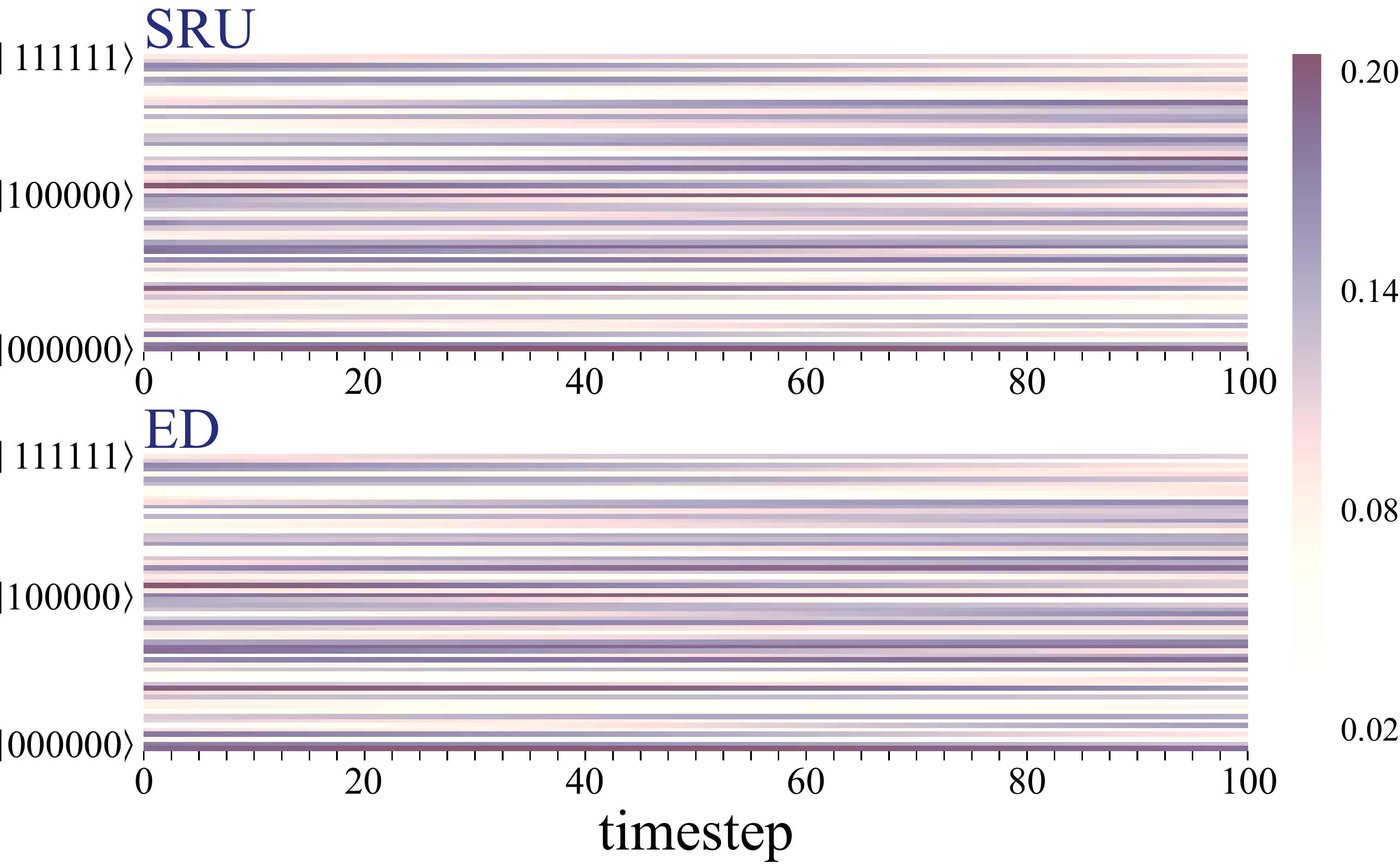}\label{6_spin_wave}}
\qquad
\subfloat[]{\includegraphics[width= .9\linewidth]{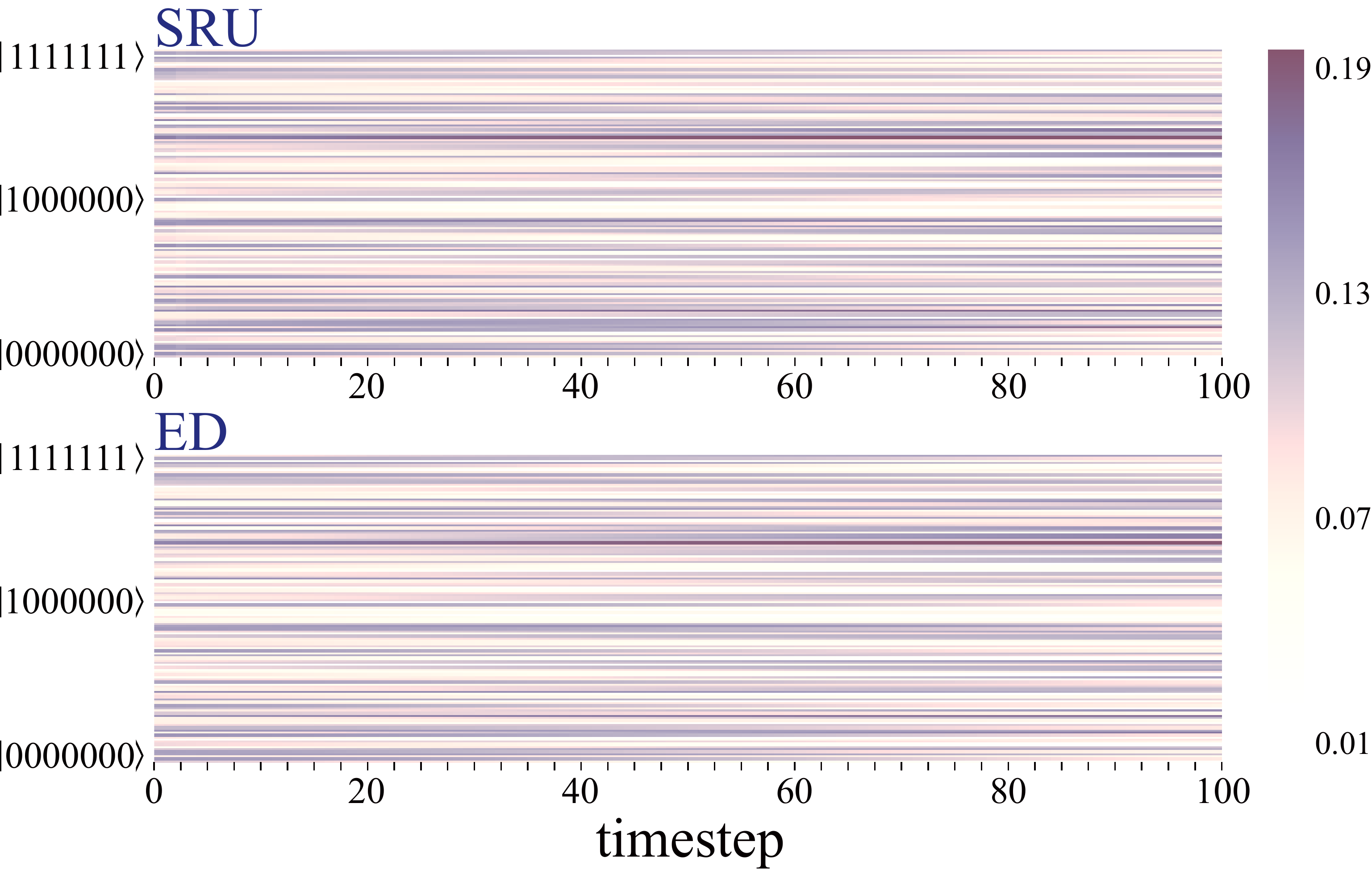}\label{7_spin_wave}}
\caption{Comparisons of all components of coefficients for SRU-based prediction and ED-based simulation in each transverse lattice with different colors for both \protect\subref{6_spin_wave} six-spin and \protect\subref{7_spin_wave} seven-spin systems.}
\label{fig:wave_function}
\end{figure}

As a concrete demonstration, we visually illustrate the comparison for the evolution of a typical state from 2-spin to 7-spin in Fig.~\ref{figresult}. These states are evolved from arbitrarily chosen initial states from the test set. Saliently, the evolution predicted by the SRU-based model accurately reproduces that from exact computations at the beginning $100$ timesteps. To confirm that this agreement is not just due to a fortuitous choice of component, we look at the evolution across all components of the same states in Fig.~\ref{fig:wave_function}. 

To further quantify the agreement of SRU and ED wavefunction evolutions, we compute the relative entropy (Kullback–Leibler divergence)\citep{vedral2002role} of their distributions over 1000 test wavefunctions sequences. For discrete probability distributions $P$ and $Q$, the relative entropy is defined as
\begin{eqnarray}
D_{KL}(P||Q)=\sum_{x}P(x)\log(\frac{P(x)}{Q(x)}).
\end{eqnarray}
Given ED-computed wavefunction coefficient vectors $M^{\textrm{ED}}$ and SRU-predicted coefficient vectors $M^{\textrm{SRU}}$, the $P$ and $Q$ variables take values
\begin{eqnarray}
P_{n,t,x}=\frac{|M^{\textrm{ED}}_{n,t,x}|}{\sum\limits_{x=1}^{2^N}|M^{\textrm{ED}}_{n,t,x}|} \\
Q_{n,t,x}=\frac{|M^{\textrm{SRU}}_{n,t,x}|}{\sum\limits_{x=1}^{2^N}|M^{\textrm{SRU}}_{n,t,x}|}
\end{eqnarray}
at time $t$ and basis vector $x$, where $n$ labels the test sequence. Hence the mean relative entropy (MRE) at each timestep $t$ is
\begin{eqnarray}
D_{KL}(P||Q)(t)=\frac{1}{1000}\sum\limits_{n=1}^{1000}\sum\limits_{x=1}^{2^N} P_{n,t,x}\log\frac{P_{n,t,x}}{Q_{n,t,x}},
\end{eqnarray}
and measures the amount of information lost when the distribution $Q$ from SRU predictions is used to represent the distribution $P$ from ED results. The smaller the value of $D_{KL}(P||Q)(t)$, the more accurate is their agreement.

\begin{figure}
\includegraphics[width=1.0 \linewidth]{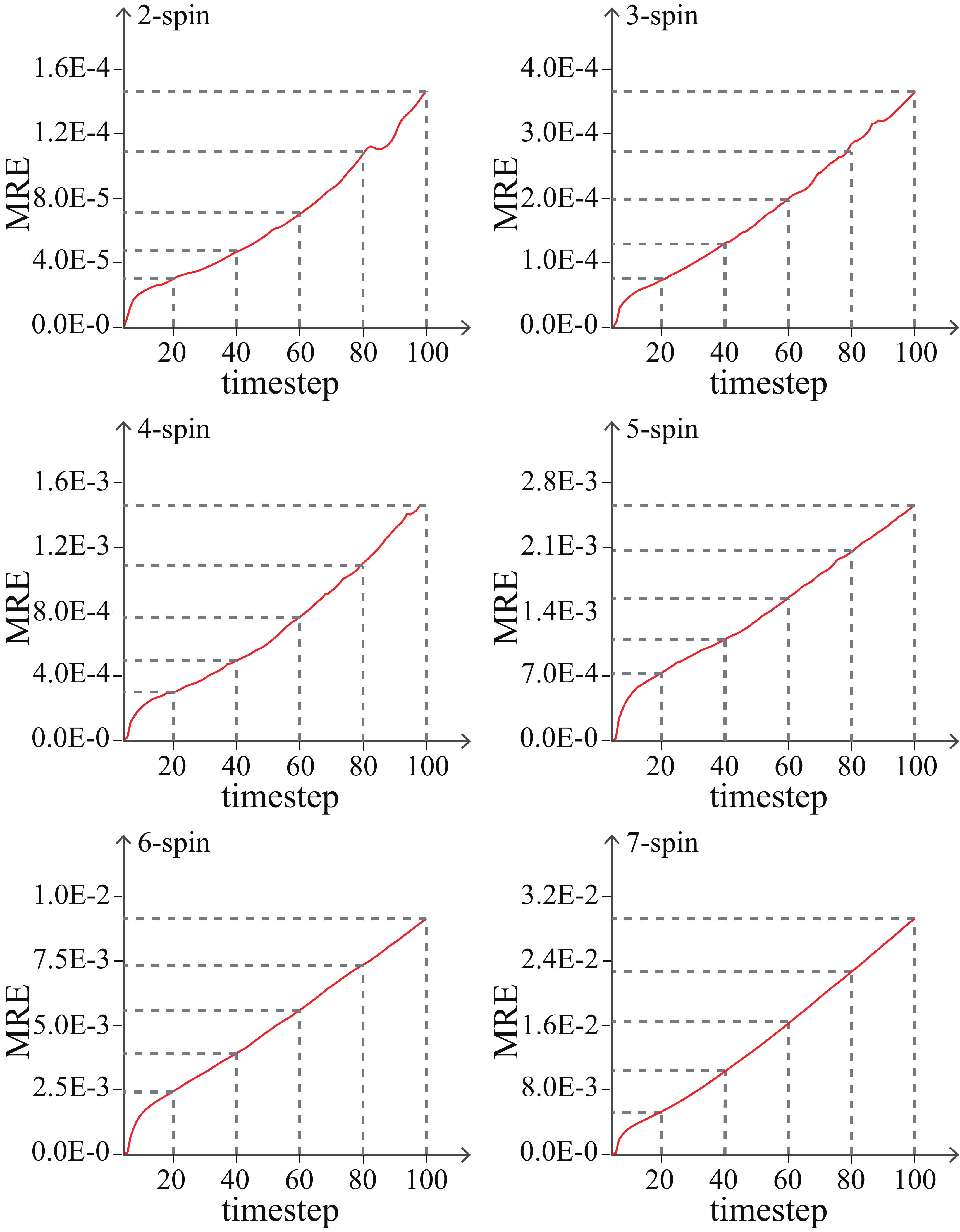}
\caption{MRE of generating long sequences in different systems. The MRE increases linearly with timesteps.}
\label{fig:kld}
\end{figure}

\begin{figure}
\includegraphics[width=1.0 \linewidth]{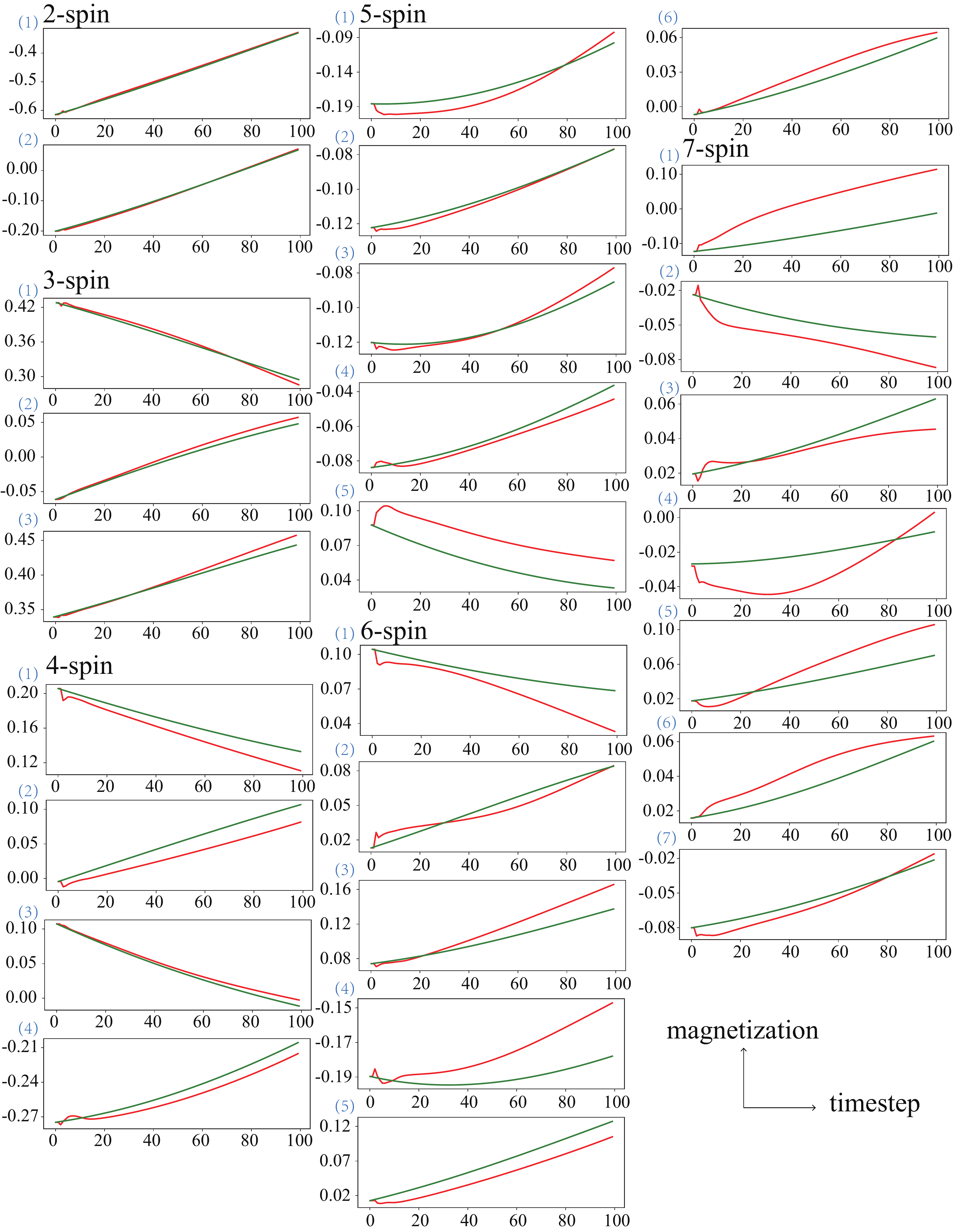}
\caption{Magnetization varies with timestep calculated by both SRU (red curve) and ED (green curve) based wavefunctions for all six systems.}
\label{fig:magnetic}
\end{figure}

In Fig.~\ref{fig:kld}, we show how the MRE varies with time during the generation of test sequences. We find that in all six systems, the order of relative entropy is always within $0.03$. Evidently, with the increase of timesteps, the relative entropy generally shows an upward trend and increases linearly with timesteps (see \citep{SOM}), which is caused by the accumulation of errors in the process of conditional generation without any external guidance, though already suppressed dropout layers. To quantify our model's performance by a physical variable, we draw the magnetization intensity calculated from both predicted (SRU) and simulated (ED) wavefunctions in Fig.~\ref{fig:magnetic}, which have a nice agreement. Specifically, for smaller-sized systems, such as 2-spin and 3-spin, the predicted magnetization intensity has a very nice agreement with simulated one. With the increase of spin variables, our SRU-based framework has a performance drop due to the exponentially increased computation complexity. Meanwhile, with the increase of timesteps, the difference between predicted and simulated magnetization intensity also becomes larger, which is due to the error accumulation during the autoregressive generation without any external guidance.

\begin{table*}
\caption{Comparsion of time consumption (seconds) between ED and SRU-based methods. Three independent runs are performed to generate sequences of different batch sizes and the average time consumption is reported. $BS$ denotes the batch size, and the bold black font indicates that our SRU-based method is superior to ED with increasing spin number and batch size.}

\begin{ruledtabular}
\begin{tabular}{ccccccccc}
 &\multicolumn{2}{c}{$BS=1$}&\multicolumn{2}{c}{$BS=64$}&\multicolumn{2}{c}{$BS=128$}&\multicolumn{2}{c}{$BS=256$}\\
 System&ED&Ours&ED&Ours&ED&Ours&ED&Ours \\
 \hline
 2-spin&0.015&$0.425$&1.1&$\mathbf{0.74}$&2.3&$\mathbf{0.83}$&$4.6$&$\mathbf{0.69}$ \\
 3-spin&0.035&$0.425$&2.2&$\mathbf{0.74}$&4.4&$\mathbf{0.69}$&$8.7$&$\mathbf{0.71}$ \\
 4-spin&0.059&$0.425$&3.8&$\mathbf{0.77}$&7.6&$\mathbf{0.66}$&$15.1$&$\mathbf{0.72}$ \\
 5-spin&0.271&$0.425$&17.5&$\mathbf{0.82}$&34.9&$\mathbf{0.76}$&$69.7$&$\mathbf{0.79}$ \\
 6-spin&0.556&$\mathbf{0.425}$&35.2&$\mathbf{0.71}$&70.5&$\mathbf{0.66}$&$141.4$&$\mathbf{0.98}$ \\
 7-spin&1.15&$\mathbf{0.425}$&73.1&$\mathbf{0.80}$&146.3&$\mathbf{0.91}$&$292.5$&$\mathbf{2.18}$ \\
\end{tabular}
\end{ruledtabular}
\label{tab:table1}
\end{table*}

Owing to its unified encoding and parallelism, our SRU-based NN is becoming increasingly more superior over the ED method in terms of efficiency, as the number of spins and batch size increase. Table~\ref{tab:table1} summarizes the results. When the number of spins gets larger, e.g. $6$ and $7$, the advantage of our SRU-based framework on inference speed becomes more and more obvious, that, is attributed to its constant computational complexity. In addition, when we enlarge the batch size to $256$ for $7$ spins, our model demonstrates a speed $130$ times faster than the ED-based method.

\begin{figure}
\includegraphics[width= 1.0\linewidth]{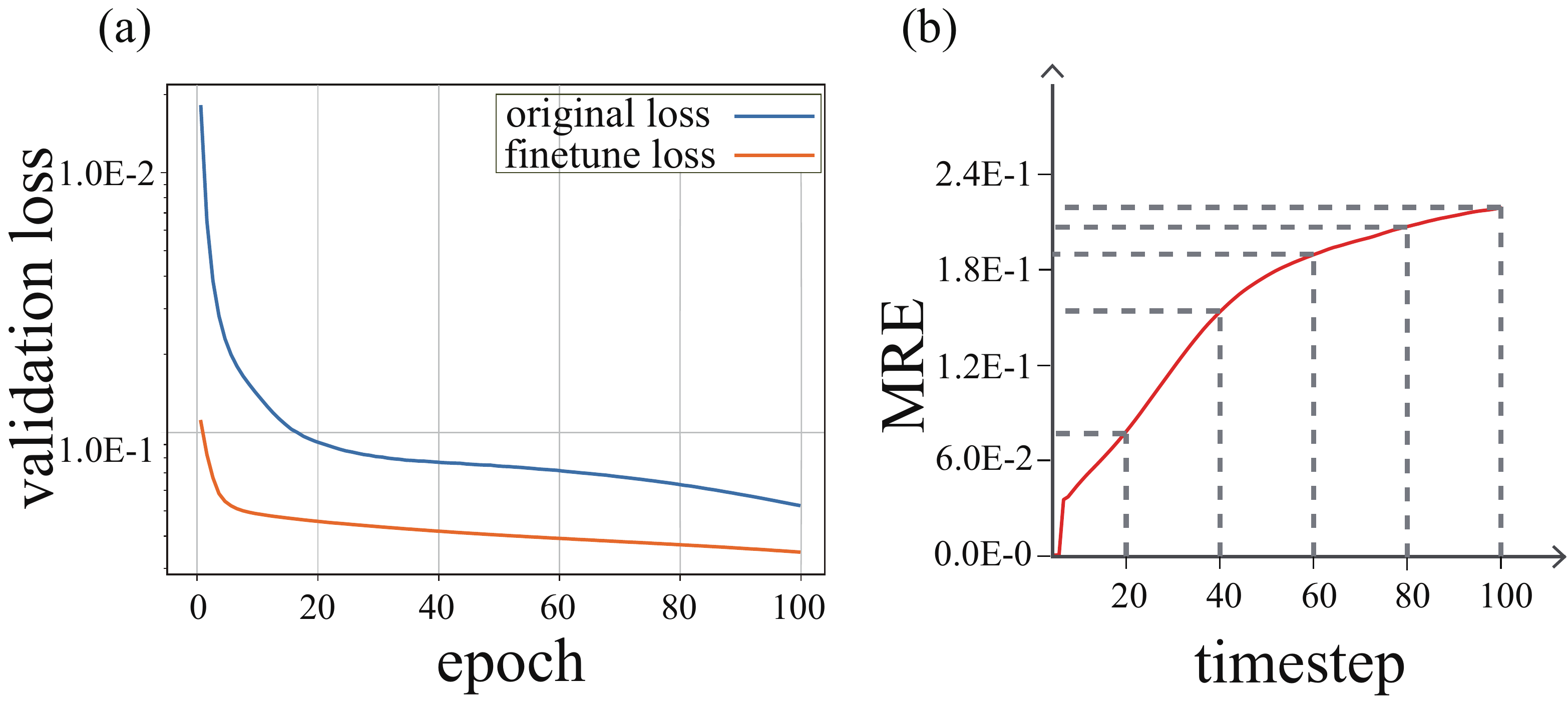}
\caption{(a) The validation loss for predicting the dynamics of the $8$-spin system by finetuning and training from scratch, respectively. (b) The MRE of $8$-spin system.}
\label{fig_loss_mre}
\end{figure}

After obtaining base model trained with datasets of $2$ to $7$ spins by $300$ epoches, we may continue to finetune it with the dataset of the $8$-spin system. To make a comparison, we also train it from scratch. The results in Fig.~\ref{fig_loss_mre}(a) shows that the validation loss by finetuning base model is much lower than training from scratch, demonstrating that our NN has already learned transferable features from smaller systems. The MRE of 8-spin system is shown in Fig.~\ref{fig_loss_mre}(b).

\section{Conclusion}
In this work, we have successfully applied a transferable NN approach based on SRU networks to approximate the state evolution of dynamic quantum many-body systems with high accuracy and superior scalability. Our work encourages future applications of advanced ML methods in quantum many-body dynamics in a Hamiltonian-agnostic manner. One possibility is to predict the behavior of large and inhomogeneous systems lack of traing data by just learning from a smaller-sized system[33]. Applications of these advancements in ML to quantum many-body problems are left to future work.

\section*{Acknowledgements}
 
Xiao Zhang thanks Yingfei Gu, Meng Cheng, Yi Zhang for discussions. Xiao Zhang is supported by the National Natural Science Foundation of China (Grant No. 11874431), the National Key R \& D Program of China (Grant No. 2018YFA0306800) and the Guangdong Science and Technology Innovation Youth Talent Program (Grant No. 2016TQ03X688). Shuo Yang is supported by NSFC (Grant No. 11804181), the National Key R \& D Program of China (Grant No. 2018YFA0306504) and the Research Fund Program of the State Key Laboratory of Low-Dimensional Quantum Physics (Grant No. ZZ201803).
\bibliography{references}

\end{document}


\begin{center}
\textbf{\large Supplemental Online Material for ``Predicting quantum many-body dynamics with transferable neural networks'' }\\[5pt]
\vspace{0.1cm}
\begin{quote}
{\small In this supplementary material, we detail: 1) the principle behind spin and chain encoding layers, SRU cell and spin decoding layer. 2) the training and inference procedure of our SRU-based NNs. 3) the initial states for illustration in our paper.}\\[20pt]
\end{quote}
\end{center}

\appendix

\renewcommand\thefigure{\thesection \arabic{figure}}
\renewcommand\thetable{\thesection \arabic{table}}
\setcounter{figure}{0}

\section{Details of SRU-based NNs}
Our proposed SRU-based ML framework is mainly composed of spin and chain encoding layers, SRU cell and spin decoding layer, we present the detailed principles behind them as follows.

\subsection{Spin and chain encoding mechanisms}
\begin{figure}
\includegraphics[width=1.0 \linewidth]{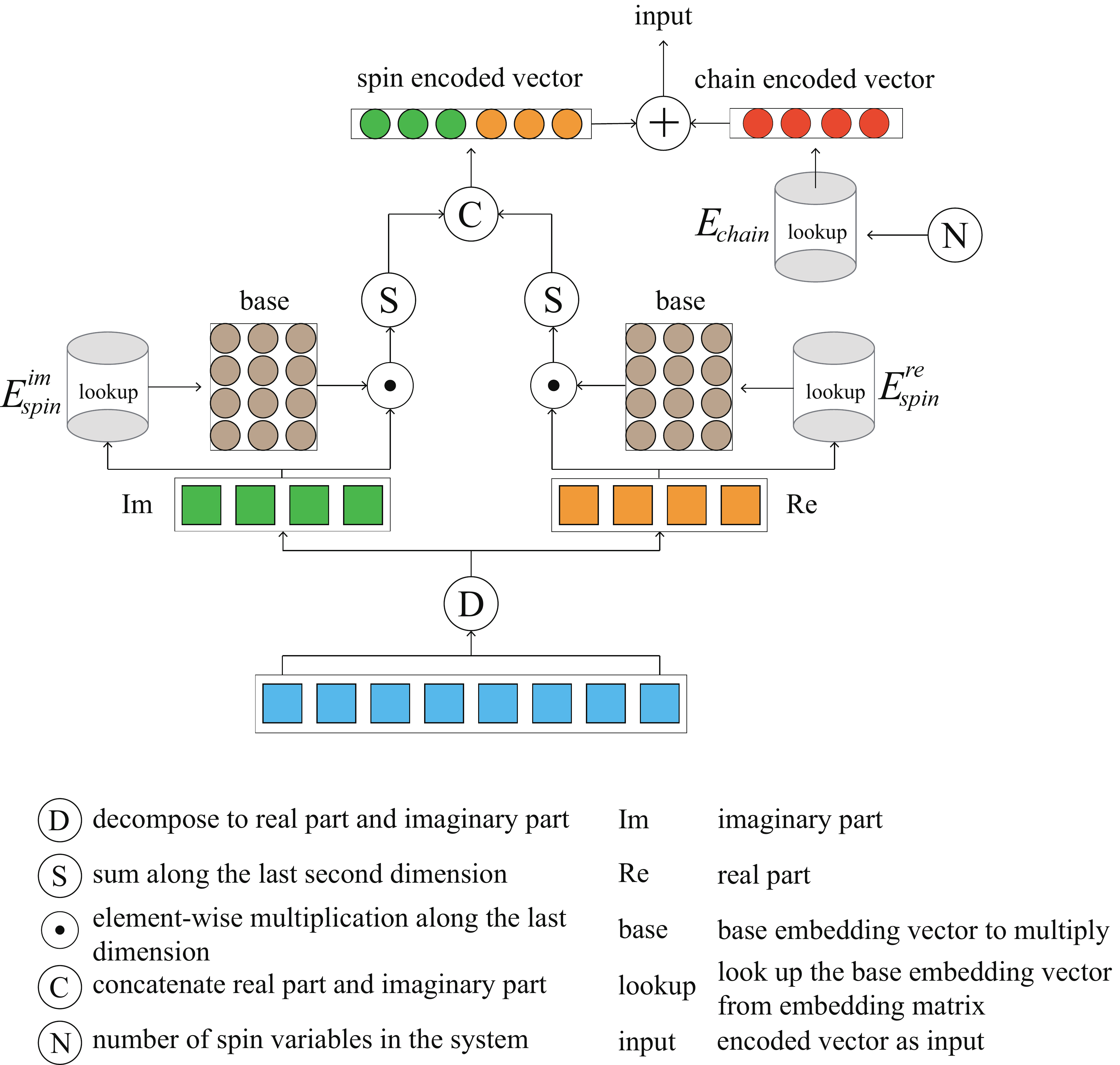}
\caption{Detailed calculation of spin and chain encoding mechanisms. The spin and chain encoded vector are concatenated into the final input vector for feed-forward layers.}
\label{fig:spin_encoding}
\end{figure}

As a prelude to the unified framework, we proposed two universal encoding mechanisms for any wavefunctions of 1D Ising spin chain, they are spin encoding and chain encoding. Both encodings are actually implemented as feed-forward neural networks, the weights of them are initially sampled from the normal distribution~\citep{glorot2010understanding}. We present the details of spin encoding and chain encoding in Fig.~\ref{fig:spin_encoding}.

The input of our SRU-based framework is composed of spin encoded vector and chain encoded vector. First, we randomly initialize two spin encoding embedding matrices $E_{spin}^{re},E_{spin}^{im}$ $\in \mathbb{R}^{2^n \times m}$ from normal distribution, here $n$ is the maximum size of spin variables, and $m$ is the hidden units for representing information of spin feature. ``re'' denotes the embedding matrix of real part and ``im'' denotes embedding matrix of the imaginary part. In this paper, we set the $n$ to be 10 (Note that $2^{10}=1024$) and the $m$ to be 256. Specifically, taking 7-spin system as example, a state sequence of its $T$ timesteps can be represented as a complex matrix $S$ $\in \mathbb{R}^{T\times {2^7}}$. We split the complex matrix $S$ into its real part $S^{re}$ and its imaginary part $S^{im}$. Since $2^7=128$, we fetch the beginning $128$ rows from $E_{spin}^{re}$ and $E_{spin}^{im}$, and then multiply them by $S^{re}$ and $S^{im}$, respectively. We take such process as ``lookup'' shown in Fig.~\ref{fig:spin_encoding}. Finally, we concatenate the real part encoded vector and imaginary part encoded vector to form the spin encoded vector ($\mathbb{R}^{T \times 512}$).  

Similarly, we randomly initialize a chain embedding matrix $E_{chain}$ $\in \mathbb{R}^{n \times m}$, where $n$ (=$10$) denotes the maximum size of spin variables of a chain, and $m$ (=$512$) is the hidden unit for storing chain feature. Specifically, taking 7-spin system as example, we adopt one-hot vector $[0,0,0,0,1,0,0,0,0,0]$ to represent such system, and multiply it by chain embedding matrix $E_{chain}$ to obtain the chain encoded vector of $\in \mathbb{R}^{T \times 512}$. Obtaining the spin encoded vector ($\mathbb{R}^{T \times 512}$) and chain encoded vector ($\mathbb{R}^{T \times 512}$), we take the sum of them as the input to SRU-based framework at each timestep. The workflow of spin and chain encoding is illustrated in Fig.~\ref{fig:spin_encoding}. 

\subsection{SRU cell}
In most RNN architectures, including Long Short-term Memory (LSTM)\citep{gers1999learning}, the new state of each timestep must be suspended until completing execution of both input and previous state. In contrast, comparing with the mainstream recurrent neural networks (RNNs), each SRU cell having a cell state and a forget gate is advantagous in terms of its efficient and parallelizable element-wise matrix product. The SRU cell is very crucial for predicting dynamics of 1D Ising model, and we assume that it can be also applied to other sequential physical tasks. 

A single layer of SRU employs the following computations:
\begin{eqnarray}
\begin{split}
        \mathbf{f}_t &= \sigma (\mathbf{W}_f \mathbf{x}_t+\mathbf{v}_f\odot \mathbf{c}_{t-1}+\mathbf{b}_f) \\
        \mathbf{c}_t &= \mathbf{f}_t \odot \mathbf{c}_{t-1}+(1-\mathbf{f}_t)\odot(\mathbf{W}\mathbf{x}_{t}) \\
        \mathbf{r}_t &= \sigma(\mathbf{W}_r \mathbf{x}_t + \mathbf{v}_r \odot \mathbf{c}_{t-1}+\mathbf{b}_r) \\
\mathbf{h}_t &= \mathbf{r}_t\odot \mathbf{c}_t+(1-\mathbf{r}_t)\odot \mathbf{x}_t
\end{split}
\label{eq:SRU}
\end{eqnarray}
\noindent where $\mathbf{f}_t$ is a forget gate that can discard the redundant information several timesteps ago and unnecessary information of old spin chains in fusion training. It is computed from a sigmoid function ($\sigma$, $\sigma(x)=\frac{1}{1+e ^{-x}}$) activated sum of input $\mathbf{x}_t$  weighted by matrix $\mathbf{W}_f$, the previous cell state $\mathbf{c}_{t-1}$ weighted by vector $\mathbf{v}_f$ and a bias vector $\mathbf{b}_f$. $\mathbf{c}_t$ is the cell state that serves to give feedbacks of latent correlated features from previous to next recursively, as the moving average of previous cell state $\mathbf{c}_t$ weighted by $\mathbf{f}_t$ plus the weighted input $\mathbf{W} \mathbf{x}_t$. $\mathbf{c}_t$ is crucial for the accurate prediction of dynamics based on just one initial quantum state. $\mathbf{r}_t$ is the skip gate of highway network \citep{srivastava2015highway} that is a shortcut for information flow with learnable weights to determine the extent of the information skip, and is computed from a sigmoid function activated sum of weighted input $\mathbf{W}_r\mathbf{x}_t$, weighted previous cell state $\mathbf{v}_r\mathbf{c}_{t-1}$ and a bias vector $\mathbf{b}_r$. The skip gate $\mathbf{r}_t$ adaptively averages the input $\mathbf{x}_t$ and the cell state $\mathbf{c}_t$ produced from the light recurrent connection. $(1-\mathbf{r}_t)\odot \mathbf{x}_t$ is a residual connection to settle the gradient dimishing problem, and the useage of highway network has been proved to improve the stability. Four SRU layers are stacked to learn better context-dependent latent information and transferable features.
\begin{algorithm}[H]
\begin{algorithmic}[1]

    \caption{Learning quantum many-body dynamics by SRU-based framework}
    \REQUIRE {$\theta$: the parameters of whole network composed of spin and chain encoding, SRU and spin decoding layers.}
    \INPUT {$B_S$ coefficient sequences in a batch with $T_L$ timesteps each.}
    \OUTPUT {same as input but with one timestep offset.}
    \PREPARATION {concatenate the real part $x_r$ and imaginary part $x_{i}$ of each complex coefficient into $x$.} 
    \item[]
    \TRAINING
    \STATE \textbf{Initialize} $\theta$ 
      \FOR {each training epoch}
      \FOR {$k$ steps}
      \STATE number of particles is $N$
	  \STATE obtain the output of spin encoding for both real part and imaginary part:
	  \STATE \indent$feature_{SE}=[Emb_{SE}(x_r)x_r;Emb_{SE}(x_i)x_i]$
	  \STATE obtain the output of chain encoding:
	  \STATE \indent$feature_{CE}=Emb_{CE}(N)$
      \STATE get the sum as input:
      \STATE \indent$input=feature_{SE}+feature_{CE}$
      \STATE pass through two feed-forward layers, apply dropout and layer normalization (LN):
      \STATE \indent$output_{f1}=LN(dropout(ReLU(W_{1}input))W_{2}+input)$
	  \STATE pass through block of four stacked SRU layers:
	  \STATE \indent$output_{s}=SRU_{block}(output_{f1})$
      \STATE pass through two feed-forward layers, apply dropout and layer normalization (LN):
      \STATE \indent$output_{f2}=LN(dropout(ReLU(W_{1}output_{s}))W_{2}+output_{s})$
	  \STATE obtain the real part of output coefficient from spin decoding:
	  \STATE \indent$c_{r}={Emb}_{SD}^{R} output_{f2}$
	  \STATE obtain the imaginary part of output coefficient from spin decoding:
	  \STATE \indent$c_{i}={Emb}_{SD}^{I} output_{f2}$
	  \STATE Update $\theta$ by ascending the stochastic gradient descent:
	  \STATE $\nabla (\theta)\left\{ \dfrac{1}{T_L}\sum\limits_{T_L}\dfrac{1}{B_S}\sum\limits_{B_S}\big[ (x_r-c_r)^2 + (x_i-c_i)^2\big]\right\}$
     \ENDFOR
     \ENDFOR
    \item[]
    \INFERENCE
      \FOR {$g$ steps}
	\IF {$g == 1$}
	  \STATE compute the first predicted coefficient vector $c_{1}$ by feeding with initial coefficient sequence as $input$: 
	  \STATE \indent$c_{1}=framework(input; \theta)$
	\ELSE
	  \STATE compute new coefficients $c_{g}$ by feeding with previous output $c_{g-1}$: 
	  \STATE \indent$c_{g}=framework(c_{g-1}; \theta)$
	\ENDIF
     \ENDFOR
    \label{alg:whole}
  \end{algorithmic}
\end{algorithm}

An obvious difference between SRU and common RNN architectures is the way of $\mathbf{c}_{t-1}$ used in the sigmoid gate. Generally, $c_{t-1}$ is involved in matrix multiplication with weight matrix to compute each gate. For example, the forget gate of a commonly adopted LSTM architecture is computed as $\mathbf{f}_t = \sigma(\mathbf{W}_{xf}x_t+\mathbf{W}_{hf}h_{t-1}+\mathbf{W}_{cf}\mathbf{c}_{t-1}+\mathbf{b}_f)$. Obviously, the matrix multiplication between $\mathbf{c}_{t-1}$ and $\mathbf{W}_{cf}$ makes the parallelization of computing state $\mathbf{f}_t$ almost impossible, because each dimension of state $\mathbf{f}_t$ can't be derived until the whole dimensions of $\mathbf{c}_{t-1}$ are available. In Eq.~(\ref{eq:SRU}), we substitute all matrix multiplications associated with $\mathbf{c}_{t-1}$ with element-wise multiplications (denoted as $\odot$). Based on this simplication, we make each state of SRU independent and parallelizable, and it is efficient and lossless.

To make SRU work efficiently, we follow the guidances of parallelized implementation performed in the context of CUDA programming~\citep{lei2018simple}. Eq.~(\ref{eq:SRU}) has totally three weight matrices: $W$, $W_f$ and $W_r$, and all of them need to be multiplied by the input sequence $\left\{ x_1\dots x_L\right\}$, where $L$ is the sequence length. To fully utilize the computational intensity of GPU, we do the batched matrix multiplications across all timesteps as:
\begin{equation}
        U^T=\begin{pmatrix}
                W \\
                W_f \\
                W_r \\
        \end{pmatrix}
        \left[x_1,x_2,\dots,x_L\right],
        \label{eq:batched_multiply}
\end{equation}
$U$ is the computed matrix. If the hidden unit of weight is $d$ in demension and the input is a mini-batch of $BS$ sequences, then $U$ would be a tensor of size $\left[L, BS, 3d\right]$.

Kernel fusion\citep{wang2010kernel,filipovivc2015optimizing} is an optimization method to reduce overhead of data transfer from global memory by fusing some sequential kernels into a single large one, to improve performance and memory locality. Specifically, we compile all element-wise multiplications into a single fused CUDA kernel function and parallelize the computation across each dimension of cell state. The complexity of each SRU layer is $\bigO (L\cdot B \cdot d)$ whereas the complexity of a vanilla LSTM layer is $\bigO (L\cdot B \cdot d^2)$.

\subsection{Spin decoding mechanism}
\begin{figure}
\includegraphics[width=1.0 \linewidth]{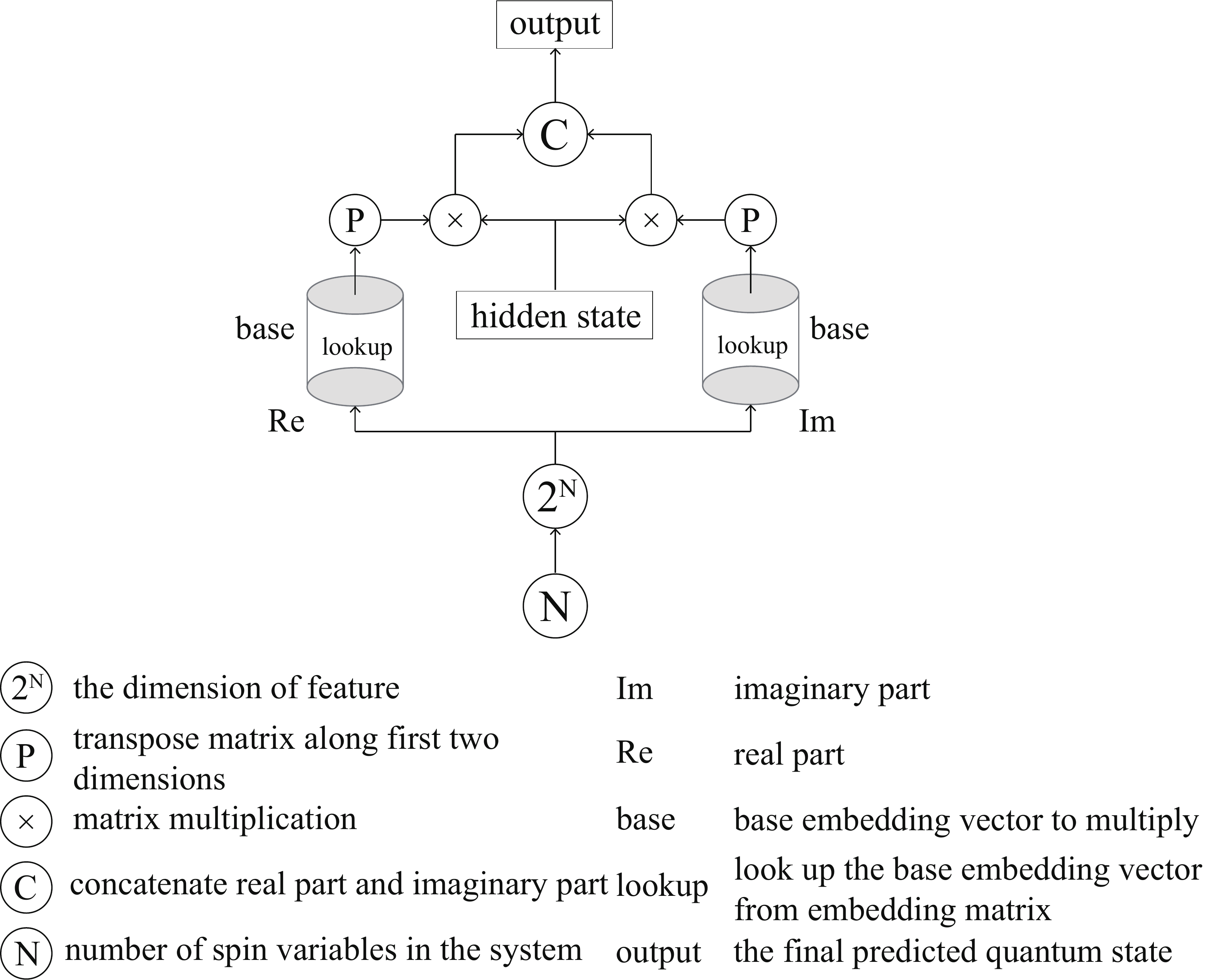}
\caption{Detailed calculation of spin decoding mechanism. The hidden state is multiplied by spin decoding embedding to form the final predicted quantum state.}
\label{fig:spin_decoding}
\end{figure}

When the hidden states of sequential modeling are obtained, we need to restore the actual predicted states. Correspondingly, we propose the spin decoding mechanism (detailed in Fig.~\ref{fig:spin_decoding}). To get the final predicted output, we need the explicit size of spin variable $N$, and then look up two embedding matrices of feature dimension $2^N$. We then transpose each embedding vector and then do matrix multiplication between embedding vector and hidden state. Finally, we concatenate the real part and imaginary part to get the predicted state. The embedding matrix $E_{dec}$$\in \mathbb{R}^{1024 \times 512}$ of spin decoding layer is also randomly initialized from normal distribution.

The overall methodology for the design of our SRU-based framework follows the criteria of unity, efficiency and transferability. In addition to the spin encoding, chain encoding, SRU module and spin decoding described above, the feed-forward layers are the bottleneck layers containing full connections with a $tanh$ nonlinearity and dropout\citep{srivastava2014dropout}, whose purpose is to distill the high-level latent feature and enable the framework to generalize well to unseen input. Such a gap between training and inference as context mismatch, mechanisms to alleviate such effect in sequence models are scheduled sampling\citep{bengio2015scheduled}, professor training\citep{lamb2016professor} and generative adversial network (GAN)\citep{goodfellow2014generative,guo2019new}. We empirically observe that just adding dropout layers also yields better scalability in the context of 1D Ising spin chain. The trainable weights and biases are initialized with zero-mean and unit-variance random distributions throughout the entire framework.  Each input state (the real and imaginary parts are concatenated together) is represented by the spin encoding layer and chain encoding layer, and then concatenated into two stacked feed-forward layers with $tanh$ activation. The dropout layer is also applied to the inputs prior to the second feed-forward layer for regularization. To be specific, we set the dropout rate to be $0.005$, that is to say, during training, about 0.5\% of neurons are set to be zero randomly. The SRU module, taking these hidden features from the dropout layer as input to construct the equivalent contextual level representation, updates the cell state and output the hidden state. Finally, a feed-forward layer with the spin decoding generates a new physical state.

\begin{figure}
\includegraphics[width=1.0 \linewidth]{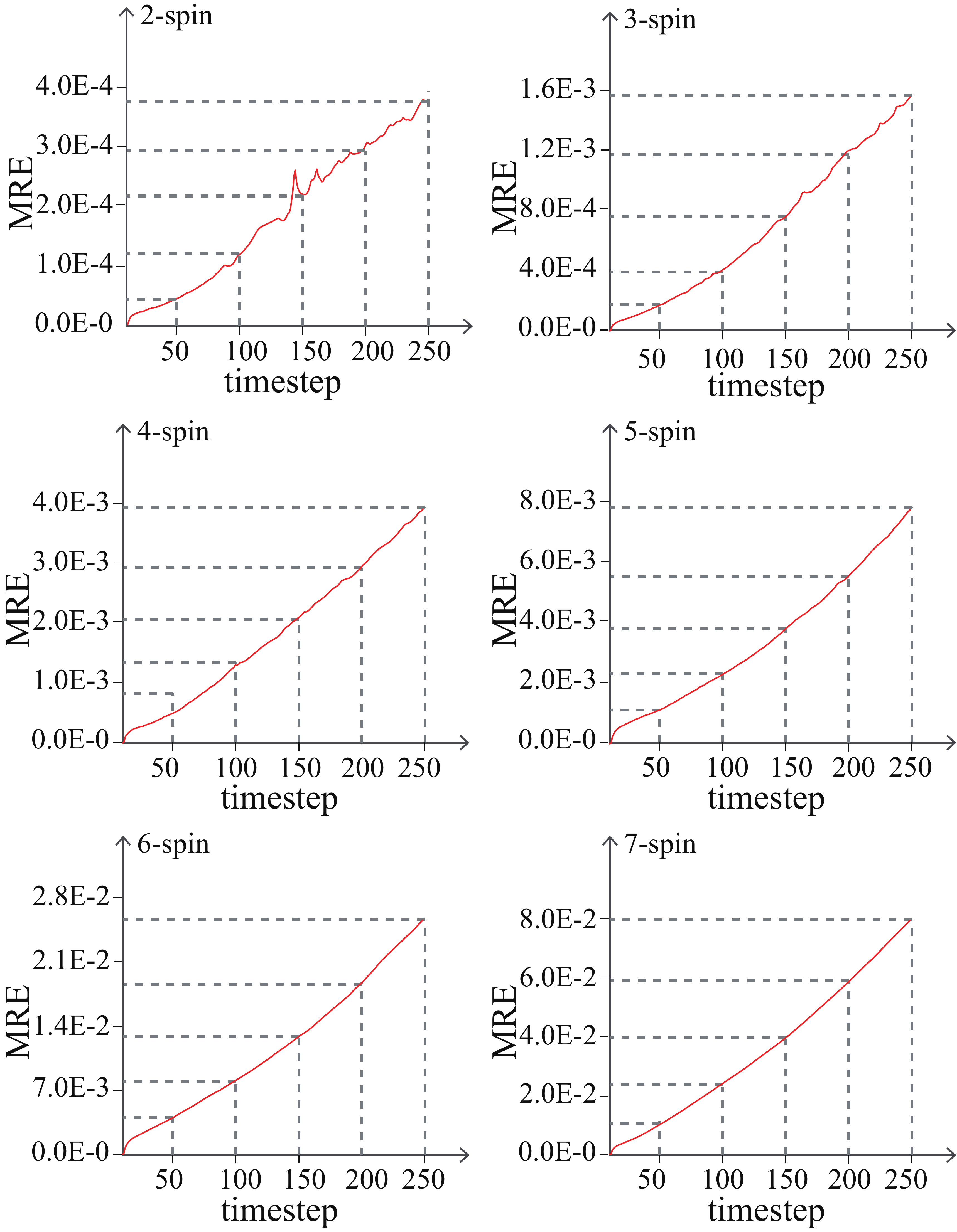}
\caption{MRE of generating longer sequences of 250 timesteps in different systems. The MRE increases linearly with timesteps.}
\label{fig:kld_250}
\end{figure}

\section{Training and inference procedure}
At the stage of training (Fig.1(c)), given a state sequence of $n$ timesteps $\{\Psi_{1},...,\Psi_{n}\}$, we can split it into two subsequences: $seq_{i}=\{\Psi_{1},\Psi_{2},...,\Psi_{n-2},\Psi_{n-1}\}$ and $seq_{o}=\{\Psi_{2},\Psi_{3}...,\Psi_{n-1},\Psi_{n}\}$. The state sequences are divided into training, validation and test sets, with an 80\%-10\%-10\% split. Considering the different scales in different systems, we observed no empirical benefits from introducing data normalization pipeline. It therefore makes sense to use raw data generated from exact diagnolization (ED) method and avoid the need of data normalization and denormalization. We use a sequence learning method which passes the input $seq_{i}$ into the network at each timestep, and computes output $seq_{o}$ of the final layer based on a linear activation function. In contrast, at the stage of inference (Fig.1(d)), we only pass the initial wavefunction into the network as input, and the network then autoregressively generates the following states at future timesteps, by consuming the previously generated states as additional input. The full procedure of training and inference is detailed in Algorithm \ref{alg:whole}.

To be more explicit, our dataset is gathered by computing the time evolution of wavefunctions from the two to seven-spin systems every $0.002$ second, using ED method\cite{Moessner2000Two}. Treated as a standard ML process, we divide the whole dataset into three parts: a training set with $10,000$ sequences, a validation set with $1,000$ sequences and a test set with $1,000$ sequences. During training, we use the mini-batch gradient descent framework to optimize our network and $m=16$ is the batch size ($B_S$) we use, considering both NN's converging speed and GPU memory. 

\begin{table}
\caption{\label{tab:parameters}Network architecture hyper-parameters. }
\begin{ruledtabular}
\begin{tabular}{cc}
layer&configuration\\
\hline
spin encoding embedding & size=1024, unit=512 \\
chain encoding embedding & size=1024, unit=512 \\
feed-forward layer & unit=512 \\
dropout layer & dropout rate=0.005 \\
feed-forward layer & unit=512 \\
SRU layer & layers=4, unit=512 \\
feed-forward layer & unit=512 \\
dropout layer & dropout rate=0.005 \\
spin decoding embedding & size=1024, unit=512 \\
batch size & 32 \\
learning rate (lr) & initial lr=0.001, decay rate=0.95
\end{tabular}
\end{ruledtabular}
\end{table}

The target of training is to minimize the mean squared error (MSE) at all timesteps between output and target states. Before training, we initialize all weight variables based on normal distribution and set all bias variables to be zero vectors. We use the adaptive moment estimation (Adam) optimizer~\citep{kingma2014adam} to update the parameters for that it's very suitable for tasks with datasets of large dimension and needs less memory space, with an initial learning rate of $0.001$ decayed by a factor of $0.95$ every epoch until the learning rate falls below $5.0e-5$. We train our base model for $300$ epoches and keep the best model with a lowest validation loss. Our model is implemented in PyTorch~\citep{paszke2017automatic} and TensorFlow~\citep{abadi2016tensorflow} framework, and all simulations are run on single NVIDIA Tesla P40 GPU and Intel Xeon CPU E5-2680. Our SRU-based framework is predominantly run on a single GPU, whereas the conventional ED-based method is run on a CPU totally. The hyper-parameters are presented in Table~\ref{tab:parameters}. 

In the transfer learning experiment, we select the base model and start training the new dataset using the same Adam optimizer with an initial learning rate of $0.0001$ decayed by a factor of $0.97$. The batch size is $16$ and the total training epochs are $300$. 

In the main text we only showed the result of predicting state sequences of $100$ timesteps. To demonstrate the ability of generating longer state sequences for different quantum many-body systems, we showed the performance of predicting state sequences as long as 250 timesteps evaluated by mean relative entropy (MRE). From Fig.~\ref{fig:kld_250}, the MRE increases linearly with the timesteps because of the error accumulation during the autoregressive generation. We will try to apply non-autoregressive generative model with inverse autoregressive flows\citep{kingma2016improved} to better estimate the probability density of longer state sequences in our future work.

\section{Initial states used for prediction}
\subsection{The initial state of two-spin system for sequence generation}
(-0.078+0.203j)$\ket{00}$+(-0.458+0.359j)$\ket{01}$ \\
+( 0.372+0.602j)$\ket{10}$+(-0.274+0.197j)$\ket{11}$ \\
\subsection{The initial state of three-spin system for sequence generation}
(-0.127+0.134j)$\ket{000}$+(-0.012+0.357j)$\ket{001}$ \\
+( 0.218+0.271j)$\ket{010}$+(-0.082+0.158j)$\ket{011}$ \\
+( 0.101+0.217j)$\ket{100}$+(-0.175+0.499j)$\ket{101}$ \\
+(-0.253+0.274j)$\ket{110}$+(-0.123+0.442j)$\ket{111}$ \\
\subsection{The initial state of four-spin system for sequence generation}
(-0.091+0.008j)$\ket{0000}$+(-0.141+0.123j)$\ket{0001}$ \\
+( 0.002+0.382j)$\ket{0010}$+( 0.169+0.190j)$\ket{0011}$ \\
+( 0.015+0.153j)$\ket{0100}$+( 0.042+0.028j)$\ket{0101}$ \\
+(-0.075+0.177j)$\ket{0110}$+( 0.137+0.255j)$\ket{0111}$ \\
+(-0.236+0.037j)$\ket{1000}$+(-0.085+0.331j)$\ket{1001}$ \\
+( 0.140+0.069j)$\ket{1010}$+(-0.230+0.270j)$\ket{1011}$ \\
+( 0.071+0.319j)$\ket{1100}$+(-0.162+0.272j)$\ket{1101}$ \\
+(-0.149+0.132j)$\ket{1110}$+(-0.016+0.169j)$\ket{1111}$ \\
\subsection{The initial state of five-spin system for sequence generation}
( 0.037+0.077j)$\ket{00000}$+(-0.110+0.026j)$\ket{00001}$ \\
+( 0.072+0.028j)$\ket{00010}$+(-0.026+0.036j)$\ket{00011}$ \\
+(-0.088+0.026j)$\ket{00100}$+(-0.102+0.268j)$\ket{00101}$ \\
+(-0.099+0.253j)$\ket{00110}$+(-0.042+0.073j)$\ket{00111}$ \\
+(-0.087+0.052j)$\ket{01000}$+( 0.055+0.120j)$\ket{01001}$ \\
+( 0.003+0.128j)$\ket{01010}$+(-0.086+0.181j)$\ket{01011}$ \\
+(-0.110+0.241j)$\ket{01100}$+( 0.110+0.050j)$\ket{01101}$ \\
+( 0.107+0.249j)$\ket{01110}$+(-0.021+0.083j)$\ket{01111}$ \\
+( 0.036+0.099j)$\ket{10000}$+( 0.119+0.257j)$\ket{10001}$ \\
+(-0.072+0.160j)$\ket{10010}$+(-0.133+0.033j)$\ket{10011}$ \\
+(-0.119+0.121j)$\ket{10100}$+(-0.113+0.213j)$\ket{10101}$ \\
+(-0.026+0.105j)$\ket{10110}$+(-0.081+0.038j)$\ket{10111}$ \\
+(-0.125+0.246j)$\ket{11000}$+( 0.027+0.185j)$\ket{11001}$ \\
+(-0.009+0.112j)$\ket{11010}$+(-0.037+0.204j)$\ket{11011}$ \\
+(-0.005+0.121j)$\ket{11100}$+(-0.017+0.015j)$\ket{11101}$ \\
+(-0.126+0.214j)$\ket{11110}$+(-0.091+0.228j)$\ket{11111}$ \\
\subsection{The initial state of six-spin system for sequence generation}
( 0.067+0.060j)$\ket{000000}$+(-0.094+0.072j)$\ket{000001}$ \\
+(-0.034+0.174j)$\ket{000010}$+(-0.067+0.151j)$\ket{000011}$ \\
+( 0.005+0.069j)$\ket{000100}$+(-0.037+0.151j)$\ket{000101}$ \\
+( 0.034+0.131j)$\ket{000110}$+( 0.023+0.038j)$\ket{000111}$ \\
+(-0.064+0.013j)$\ket{001000}$+(-0.084+0.030j)$\ket{001001}$ \\
+( 0.082+0.061j)$\ket{001010}$+( 0.051+0.146j)$\ket{001011}$ \\
+( 0.043+0.022j)$\ket{001100}$+( 0.001+0.141j)$\ket{001101}$ \\
+( 0.058+0.015j)$\ket{001110}$+( 0.065+0.014j)$\ket{001111}$ \\
+(-0.041+0.035j)$\ket{010000}$+( 0.007+0.131j)$\ket{010001}$ \\
+( 0.002+0.126j)$\ket{010010}$+( 0.025+0.158j)$\ket{010011}$ \\
+( 0.042+0.013j)$\ket{010100}$+(-0.006+0.006j)$\ket{010101}$ \\
+( 0.090+0.086j)$\ket{010110}$+( 0.093+0.022j)$\ket{010111}$ \\
+( 0.082+0.154j)$\ket{011000}$+(-0.011+0.114j)$\ket{011001}$ \\
+(-0.040+0.081j)$\ket{011010}$+(-0.004+0.107j)$\ket{011011}$ \\
+(-0.088+0.199j)$\ket{011100}$+(-0.053+0.074j)$\ket{011101}$ \\
+( 0.093+0.159j)$\ket{011110}$+(-0.088+0.119j)$\ket{011111}$ \\
+( 0.055+0.134j)$\ket{100000}$+( 0.015+0.126j)$\ket{100001}$ \\
+( 0.031+0.006j)$\ket{100010}$+( 0.069+0.094j)$\ket{100011}$ \\
+( 0.003+0.180j)$\ket{100100}$+(-0.058+0.109j)$\ket{100101}$ \\
+(-0.034+0.075j)$\ket{100110}$+( 0.071+0.022j)$\ket{100111}$ \\
+( 0.031+0.153j)$\ket{101000}$+( 0.033+0.191j)$\ket{101001}$ \\
+(-0.046+0.193j)$\ket{101010}$+( 0.025+0.013j)$\ket{101011}$ \\
+( 0.062+0.165j)$\ket{101100}$+(-0.045+0.068j)$\ket{101101}$ \\
+( 0.048+0.111j)$\ket{101110}$+( 0.001+0.065j)$\ket{101111}$ \\
+(-0.008+0.029j)$\ket{110000}$+(-0.071+0.106j)$\ket{110001}$ \\
+( 0.092+0.185j)$\ket{110010}$+(-0.066+0.075j)$\ket{110011}$ \\
+(-0.001+0.129j)$\ket{110100}$+(-0.075+0.030j)$\ket{110101}$ \\
+(-0.031+0.079j)$\ket{110110}$+( 0.005+0.056j)$\ket{110111}$ \\
+( 0.088+0.080j)$\ket{111000}$+(-0.076+0.084j)$\ket{111001}$ \\
+( 0.029+0.003j)$\ket{111010}$+(-0.072+0.074j)$\ket{111011}$ \\
+(-0.076+0.175j)$\ket{111100}$+(-0.075+0.009j)$\ket{111101}$ \\
+(-0.009+0.187j)$\ket{111110}$+( 0.032+0.199j)$\ket{111111}$ \\
\subsection{The initial state of seven-spin system for sequence generation}
( 0.067+0.007j)$\ket{0000000}$+( 0.018+0.119j)$\ket{0000001}$ \\
+(-0.013+0.033j)$\ket{0000010}$+( 0.050+0.115j)$\ket{0000011}$ \\
+(-0.039+0.098j)$\ket{0000100}$+(-0.036+0.055j)$\ket{0000101}$ \\
+(-0.020+0.011j)$\ket{0000110}$+(-0.005+0.073j)$\ket{0000111}$ \\
+(-0.004+0.125j)$\ket{0001000}$+(-0.050+0.098j)$\ket{0001001}$ \\
+( 0.010+0.111j)$\ket{0001010}$+( 0.056+0.085j)$\ket{0001011}$ \\
+(-0.008+0.092j)$\ket{0001100}$+( 0.017+0.003j)$\ket{0001101}$ \\
+( 0.057+0.012j)$\ket{0001110}$+(-0.040+0.058j)$\ket{0001111}$ \\
+( 0.021+0.053j)$\ket{0010000}$+(-0.047+0.053j)$\ket{0010001}$ \\
+( 0.057+0.111j)$\ket{0010010}$+( 0.052+0.006j)$\ket{0010011}$ \\
+( 0.010+0.044j)$\ket{0010100}$+(-0.010+0.105j)$\ket{0010101}$ \\
+( 0.052+0.128j)$\ket{0010110}$+(-0.005+0.072j)$\ket{0010111}$ \\
+(-0.034+0.119j)$\ket{0011000}$+(-0.015+0.083j)$\ket{0011001}$ \\
+( 0.065+0.100j)$\ket{0011010}$+(-0.018+0.009j)$\ket{0011011}$ \\
+( 0.019+0.072j)$\ket{0011100}$+(-0.065+0.129j)$\ket{0011101}$ \\
+( 0.035+0.013j)$\ket{0011110}$+( 0.040+0.117j)$\ket{0011111}$ \\
+( 0.011+0.114j)$\ket{0100000}$+( 0.059+0.039j)$\ket{0100001}$ \\
+( 0.057+0.035j)$\ket{0100010}$+( 0.034+0.108j)$\ket{0100011}$ \\
+( 0.066+0.131j)$\ket{0100100}$+(-0.002+0.078j)$\ket{0100101}$ \\
+( 0.009+0.108j)$\ket{0100110}$+( 0.020+0.087j)$\ket{0100111}$ \\
+( 0.010+0.069j)$\ket{0101000}$+( 0.018+0.090j)$\ket{0101001}$ \\
+(-0.052+0.070j)$\ket{0101010}$+( 0.031+0.022j)$\ket{0101011}$ \\
+( 0.001+0.106j)$\ket{0101100}$+(-0.063+0.062j)$\ket{0101101}$ \\
+(-0.049+0.057j)$\ket{0101110}$+(-0.014+0.043j)$\ket{0101111}$ \\
+(-0.048+0.029j)$\ket{0110000}$+( 0.047+0.049j)$\ket{0110001}$ \\
+( 0.002+0.097j)$\ket{0110010}$+( 0.065+0.124j)$\ket{0110011}$ \\
+( 0.026+0.034j)$\ket{0110100}$+(-0.014+0.083j)$\ket{0110101}$ \\
+( 0.021+0.080j)$\ket{0110110}$+(-0.051+0.001j)$\ket{0110111}$ \\
+(-0.052+0.044j)$\ket{0111000}$+( 0.002+0.092j)$\ket{0111001}$ \\
+( 0.053+0.050j)$\ket{0111010}$+( 0.063+0.013j)$\ket{0111011}$ \\
+( 0.023+0.082j)$\ket{0111100}$+(-0.014+0.001j)$\ket{0111101}$ \\
+( 0.011+0.123j)$\ket{0111110}$+(-0.025+0.066j)$\ket{0111111}$ \\
+(-0.038+0.019j)$\ket{1000000}$+(-0.000+0.049j)$\ket{1000001}$ \\
+(-0.034+0.018j)$\ket{1000010}$+(-0.056+0.056j)$\ket{1000011}$ \\
+(-0.028+0.047j)$\ket{1000100}$+(-0.016+0.057j)$\ket{1000101}$ \\
+( 0.032+0.094j)$\ket{1000110}$+( 0.056+0.114j)$\ket{1000111}$ \\
+(-0.006+0.132j)$\ket{1001000}$+( 0.042+0.071j)$\ket{1001001}$ \\
+( 0.011+0.036j)$\ket{1001010}$+( 0.034+0.093j)$\ket{1001011}$ \\
+( 0.040+0.047j)$\ket{1001100}$+(-0.003+0.069j)$\ket{1001101}$ \\
+(-0.022+0.075j)$\ket{1001110}$+( 0.030+0.078j)$\ket{1001111}$ \\
+(-0.014+0.049j)$\ket{1010000}$+(-0.059+0.120j)$\ket{1010001}$ \\
+( 0.024+0.050j)$\ket{1010010}$+( 0.008+0.118j)$\ket{1010011}$ \\
+(-0.045+0.047j)$\ket{1010100}$+( 0.060+0.047j)$\ket{1010101}$ \\
+( 0.030+0.089j)$\ket{1010110}$+(-0.031+0.007j)$\ket{1010111}$ \\
+( 0.066+0.051j)$\ket{1011000}$+( 0.035+0.001j)$\ket{1011001}$ \\
+(-0.010+0.055j)$\ket{1011010}$+(-0.062+0.043j)$\ket{1011011}$ \\
+( 0.039+0.015j)$\ket{1011100}$+(-0.015+0.084j)$\ket{1011101}$ \\
+( 0.018+0.124j)$\ket{1011110}$+( 0.021+0.029j)$\ket{1011111}$ \\
+( 0.037+0.061j)$\ket{1100000}$+( 0.053+0.027j)$\ket{1100001}$ \\
+( 0.059+0.127j)$\ket{1100010}$+(-0.037+0.064j)$\ket{1100011}$ \\
+(-0.041+0.072j)$\ket{1100100}$+( 0.007+0.121j)$\ket{1100101}$ \\
+(-0.054+0.045j)$\ket{1100110}$+( 0.010+0.063j)$\ket{1100111}$ \\
+(-0.041+0.122j)$\ket{1101000}$+(-0.064+0.089j)$\ket{1101001}$ \\
+( 0.037+0.053j)$\ket{1101010}$+(-0.050+0.085j)$\ket{1101011}$ \\
+( 0.067+0.120j)$\ket{1101100}$+( 0.030+0.055j)$\ket{1101101}$ \\
+(-0.007+0.063j)$\ket{1101110}$+(-0.062+0.022j)$\ket{1101111}$ \\
+( 0.032+0.077j)$\ket{1110000}$+( 0.044+0.102j)$\ket{1110001}$ \\
+( 0.067+0.116j)$\ket{1110010}$+( 0.061+0.009j)$\ket{1110011}$ \\
+(-0.062+0.122j)$\ket{1110100}$+(-0.065+0.037j)$\ket{1110101}$ \\
+(-0.015+0.024j)$\ket{1110110}$+( 0.049+0.067j)$\ket{1110111}$ \\
+(-0.027+0.082j)$\ket{1111000}$+( 0.041+0.003j)$\ket{1111001}$ \\
+( 0.062+0.074j)$\ket{1111010}$+(-0.049+0.118j)$\ket{1111011}$ \\
+( 0.034+0.121j)$\ket{1111100}$+( 0.022+0.063j)$\ket{1111101}$ \\
+(-0.008+0.132j)$\ket{1111110}$+(-0.066+0.069j)$\ket{1111111}$ \\

\bibliography{references}